\documentclass[10pt]{article}

\usepackage{amsthm,amsmath,latexsym,color,subfigure,setspace, multirow,soul}
\usepackage{url}\RequirePackage[colorlinks,citecolor=blue, linkcolor=blue,urlcolor = blue]{hyperref}
\usepackage{amssymb}

\usepackage{graphicx}
\usepackage{epstopdf}
\usepackage{pdflscape}

\usepackage{cite}
\usepackage{enumerate}

\usepackage{xr}

\usepackage[labelfont=bf,labelsep=period,justification=raggedright]{caption}

\makeatletter
\renewcommand{\@biblabel}[1]{\quad#1.}
\makeatother

\usepackage{adjustbox}
\usepackage{array}
\usepackage{relsize}
\usepackage{sidecap}
\usepackage{bm}
\usepackage{bbm}
\usepackage{multicol}
\usepackage{float}
\usepackage{booktabs}
\usepackage{multirow,multicol}

\newtheorem*{theorem*}{Theorem}

\newtheorem*{claim*}{Claim}
\theoremstyle{definition}

\theoremstyle{AppDefinition}

\theoremstyle{remark}

\def\eq#1{(\ref{#1})}

\def\beginmat{ \left( \begin{array} }
\def\endmat{ \end{array} \right) }

\def\log{{\rm log}}
\def\tr{{\rm tr}}
\def\cond{\, | \,}
\newcommand*\diff{\mathop{}\!\mathrm{d}}
\newcolumntype{P}[1]{>{\centering\arraybackslash}p{#1}}

\date{}

\newcommand{\bx}{\mathbf{x}}
\newcommand{\by}{\mathbf{y}}

\newcommand{\bK}{\mathbf{K}}

\newcommand{\bX}{\mathbf{X}}

\newcommand{\bZ}{\mathbf{Z}}

\newcommand{\bI}{\mathbf{I}}

\newcommand{\T}{\intercal}

\newcommand{\cN}{\mathcal{N}}
\newcommand{\cX}{\mathcal{X}}
\newcommand{\cGP}{\mathcal{GP}}

\newcommand{\E}{\mathbb{E}}
\newcommand{\V}{\mathbb{V}}

\newcommand{\bvarepsilon}{\boldsymbol\varepsilon}
\newcommand{\bbeta}{\boldsymbol\beta}

\newcommand{\btheta}{\boldsymbol\theta}

\newcommand{\bomega}{\boldsymbol\omega}

\newcolumntype{R}[2]{%
    >{\adjustbox{angle=#1,lap=\width-(#2)}\bgroup}%
    l%
    <{\egroup}%
}

\allowdisplaybreaks

\begin{document}
\def\spacingset#1{\renewcommand{\baselinestretch}%
{#1}\small\normalsize} \spacingset{1}
\begin{flushleft}
{\Large{\textbf{Variable Prioritization in Nonlinear Black Box Methods:\\ A Genetic Association Case Study}}}
\newline
\\
Lorin Crawford\textsuperscript{1-3,$\dagger$}, Seth R.~Flaxman\textsuperscript{4,5}, Daniel E.~Runcie\textsuperscript{6}, and Mike West\textsuperscript{7}

\medskip
\bf{1} Department of Biostatistics, Brown University, Providence, RI, USA
\\
\bf{2} Center for Statistical Sciences, Brown University, Providence, RI, USA
\\
\bf{3} Center for Computational Molecular Biology, Brown University, Providence, RI, USA
\\
\bf{4} Department of Mathematics, Imperial College London, London, UK
\\
\bf{5} Data Science Institute, Imperial College London, London, UK
\\
\bf{6} Department of Plant Sciences, University of California, Davis, CA, USA
\\
\bf{7} Department of Statistical Science, Duke University, Durham, NC, USA
\\
\medskip
$\dagger$ Corresponding Email: lorin\_crawford@brown.edu
\end{flushleft}


\section*{Abstract}

The central aim in this paper is to address variable selection questions in nonlinear and nonparametric regression. Motivated by statistical genetics, where nonlinear interactions are of particular interest, we introduce a novel and interpretable way to summarize the relative importance of predictor variables. Methodologically, we develop the ``RelATive cEntrality'' (RATE) measure to prioritize candidate genetic variants that are not just marginally important, but whose associations also stem from significant covarying relationships with other variants in the data. We illustrate RATE through Bayesian Gaussian process regression, but the methodological innovations apply to other ``black box'' methods. It is known that nonlinear models often exhibit greater predictive accuracy than linear models, particularly for phenotypes generated by complex genetic architectures. With detailed simulations and two real data association mapping studies, we show that applying RATE enables an explanation for this improved performance.


\section{Introduction}\label{sec1}

Classical statistical models and modern machine learning methodology have recently been dichotomized into two separate groups. The former are often characterized as interpretable modeling approaches and include conventional methods such as linear and logistic regressions. The latter, however, have sparked a greater debate as they have been frequently criticized as ``black box'' techniques with opaque implementations and uncertain internal workings. Whenever support vector machines or neural networks give meaningful performance gains over more conventional regression models, a challenge of interpretability arises. In these situations, it is often questioned what characteristics of the input data are being most used by the black box. One of the key features leading to these performance gains is the automatic inclusion of higher order interactions between variables \cite{Cotter:2011aa}. Popular machine learning kernel functions and fully connected neural network layers implicitly enumerate all possible nonlinear effects \cite{Wahba:1990aa}. While this fact is in itself a partial explanation for improvement gains, we often wish to know precisely which variables are the most important --- with the ultimate goals of furthering scientific understanding and performing model/feature selection \cite{Barbieri:2004aa}.

As our main contribution, we propose a ``RelATive cEntrality'' measure (RATE) for investigating variable importance in Bayesian nonlinear models, particularly those considered to be black box. Here, RATE identifies variables which are not just marginally important, but also those whose data associations stem from a significant covarying relationship with other variables. Our method is entirely general with respect to the modeling approach taken --- the only requirement being that a method can produce uncertainty intervals for predictions. As an illustration, we focus on Gaussian process modeling with Markov chain Monte Carlo (MCMC) inference. In addition, we note that this variable selection approach immediately applies to other methodologies such as Bayesian neural networks \cite{Richard:1991aa}, Bayesian additive regression trees \cite{Chipman:2010aa}, and approximate inference methods like variational Bayes \cite{Rasmussen}. 

While variable selection is the main utility for our method, we are motivated by the approach of continuous model expansion \cite{Gelman:2014aa}: the goal is to build the best fitting or optimally predictive model while searching over many variables and the interactions between them, but without explicitly worrying about sparsity. Indeed, this has become a recent focus of statistical methods research, especially in terms of understanding the relative importance of subsets of candidate predictors with respect to specific predictive goals \cite{Lin:2016aa}. While we believe strongly in regularization as a key ingredient in developing good statistical models, our choice of Gaussian process priors achieves robust inference without explicitly imposing a sparsity penalty. The reason to avoid sparsity constraints like the lasso is not just philosophical --- as typically applied L1-regularization suffers from a lack of stability \cite{lim2016estimation,piironen2017comparison}, and the use of Laplacian priors has too been criticized \cite{Carvalho:2010aa}. Simultaneously, we are also motivated by the rise of deep neural networks, which are typically wildly over-parameterized and yet, when combined with large datasets, can give quite impressive improvements to model performance.

We assess our proposed approach in the context of association mapping (i.e.~inference of significant variants or loci) in statistical genetics, as a way to highlight data science applications that are driven by many covarying and interacting predictors. For example, understanding how statistical epistasis between genes (i.e.~the polynomial terms of the variables in the genotype matrix) influence the architecture of traits and variation in phenotypes is of great interest in genetics applications \cite{Zhang:2007aa,Phillips:2008aa,Wan:2010aa,Zhang:2010aa,Prabhu:2012aa,Mackay:2014aa,Crawford:2017ab,Crawford:2018aa}. However, despite studies that have detected ``pervasive epistasis'' in many model organisms \cite{Horn:2011aa} and improved genomic selection (i.e.~phenotypic prediction) using nonlinear regression models \cite{Howard:2014aa}, substantial controversies remain \cite{Hill:2008aa}. For example, in some settings, association mapping studies have identified many candidates of statistical epistasis or interactions that contribute to quantitative traits \cite{Hemani:2014aa}, but some of these results can be explained by additive effects of other unsequenced variants \cite{Wood:2014aa}. To date, we have a limited understanding of this important biological question because it is often difficult to pinpoint how nonlinearities influence complex prioritization of associated genetic markers. Indeed, it has been suggested that if one aims to infer biological interactions, statistically modeled interactions and main effect terms should not be interpreted separately \cite{Wang:2010aa,Wang:2011aa}. Our contribution in this paper is therefore of direct scientific relevance in that RATE will enable scientists to consider embracing machine learning-type approaches by allowing them to open up the black box.

The remainder of this paper is organized as follows. In Section \ref{sec2}, we briefly detail the Gaussian process regression model and motivate the need for an effect size (regression coefficient) analog that serves to characterize the importance of the original input variables in nonparametric methods. In Section \ref{sec3}, we specify how to conduct association mapping using distributional centrality measures. Here, we also define the concept of relative centrality (RATE) which provides evidence for the relative importance of each variable. In Section \ref{sec4}, we show the utility of our methodology on real and simulated data. Finally, we close with a discussion in Section \ref{sec5}.


\section{Motivating Bayesian Nonparametric Framework}\label{sec2}

In this paper, we propose a relative centrality measure as an interpretable way to summarize the importance of input variables for nonparametric methodologies. We will do this within the context of association mapping in statistical genetics. This effort will require the utilization of three components: (i) a motivating probabilistic model, (ii) a notion of an effect size (or regression coefficient) for each genetic variant, and (iii) a statistical metric that determines marker significance. Each of these components are naturally given in linear regression. Our goal is to provide a computationally tractable way to derive the same necessary components for nonlinear methods. 

In this section, we focus on formulating components (i) and (ii), while component (iii) is developed later in Section \ref{sec3}. First, we begin by detailing Bayesian Gaussian process regression as our motivating probabilistic model. Next, we generalize a previous result which defines an effect size (regression coefficient) analog for the input data in nonparametric methods \cite{Crawford:2017aa}. Extensions to other methodologies (e.g.~Bayesian kernel ridge regression, neural networks) can be found in Supplementary Material. For simplicity, we make the assumption that the phenotypic response is continuous; although, the frameworks discussed can be altered for dichotomous traits (e.g.~case-control studies). This expansion would include steps similar to those outlined in previous works \cite{MJ:2011aa}. We leave these specific details to the reader.

\subsection{Gaussian Process Regression}

We now state a Bayesian modeling framework which we use to construct a generalized projection operator between an infinite dimensional function space, called a reproducing kernel Hilbert space (RKHS), and the original genotype space. This projection will allow us to define an effect size analog for Bayesian nonparametric analyses. We begin by considering standard linear regression
\begin{align}
\mathbf{y} = \mathbf{X}\bm{\beta}+\bm{\varepsilon}, \quad \bm{\varepsilon}\sim\cN(\bm{0},\tau^2\mathbf{I}),\label{mod1}
\end{align}
where $\mathbf{y}$ is an $n$-dimensional vector of phenotypes from $n$ individuals, $\mathbf{X}$ is an $n\times p$ matrix of genotypes for $p$ genetic variants encoded as $\{0,1,2\}$ copies of a reference allele at each marker, $\bm{\beta}$ is the corresponding additive effect size, $\bm{\varepsilon}$ is assumed to follow a multivariate normal distribution with mean zero and variance $\tau^2$, and $\mathbf{I}$ is an identity matrix. For convenience, we will also assume that the genotype vector has been centered and standardized to have mean 0 and standard deviation 1.

In genetic applications, the assumption that phenotypic variation can be fully explained by additive effects is often too restrictive \cite{Phillips:2008aa,Mackay:2014aa}. One natural way to overcome this problem is to conduct model inference within a high dimensional function space. Indeed, an RKHS may be defined based on a nonlinear transformation of the data using a positive definite covariance function (or kernel) that is assumed to have a finite integral operator with eigenfunctions $\{\phi_\ell\}_{\ell=1}^\infty$ and eigenvalues $\{\delta_\ell\}_{\ell=1}^\infty$. Namely,
\begin{align*}
\int k(\mathbf{x},\mathbf{x}^\prime)\mathrm{d}(\mathbf{x},\mathbf{x}^\prime) <\infty, \quad \delta_\ell\phi_\ell(\mathbf{x}) = \int_{\cX} k(\mathbf{x},\mathbf{x}^\prime) \phi_\ell(\mathbf{x}^\prime) \diff \mathbf{x}^\prime.
\end{align*} 
For these classes of covariance functions, the following infinite expansion holds $k(\mathbf{x},\mathbf{x}^\prime) = \sum_{\ell=1}^\infty \delta_\ell \phi_\ell(\mathbf{x})\phi_\ell(\mathbf{x}^\prime)$ \cite{Merc:1909}, and an RKHS function space may be formally defined via the closure of a linear combination of basis functions \cite{Wolpert:2007aa}. As a direct result, we rewrite equation Equation \eq{mod1} as the following RKHS regression model \cite{MJ:2011aa}
\begin{align}
\mathbf{y} = \bm{\Psi}^{\T}\mathbf{c}+\bm{\varepsilon}, \quad \bm{\varepsilon}\sim\cN(\mathbf{0},\tau^2\mathbf{I}),\label{mod2}
\end{align}
where ${\boldsymbol\psi}(\mathbf{x}) = \{\sqrt{\delta_\ell}\phi_\ell(\mathbf{x})\}_{\ell=1}^\infty$ is a vector space spanned by the bases, $\bm{\Psi}= [\bm{\psi}(\mathbf{x}_1),\ldots,\bm{\psi}(\mathbf{x}_n)]^{\T}$ is a corresponding matrix of concatenated basis functions, and $\mathbf{c} = \{\mathrm{c}_\ell\}_{\ell=1}^\infty$ are the corresponding basis coefficients. The above specification in Equation \eq{mod2} closely resembles the linear model in Equation \eq{mod1} --- except now the bases are the feature vectors $\bm{\psi}(\mathbf{x})$ (rather than the unit basis), and the transformed space can be infinite-dimensional. Theoretically, this is an important property because the inclusion of nonlinear interactions and covarying relationships are implicitly captured in the RKHS. 

Unfortunately, properly representing any given basis function in an empirically amenable form is a difficult task \cite{Scholkopf:2001aa}. To circumvent this analytical issue, one may alternatively conduct inference in an RKHS by specifying a Gaussian process (GP) as the prior distribution over the function space directly
\begin{align*}
f(\mathbf{x})\sim\cGP(m(\mathbf{x}),k(\mathbf{x},\mathbf{x}^\prime)),
\end{align*}
where $f(\bullet)$ is completely specified by its mean function and positive definite covariance (kernel) function, $m(\bullet)$ and $k(\bullet,\bullet)$, respectively. In practice, if we condition on a finite set of locations (i.e.~the set of observed samples $n$), the Gaussian process prior then becomes a multivariate normal \cite{Kolmogorov:1960aa}. By specifying a joint version of the nonparametric regression model above, we consider taking a ``weight-space'' view on Gaussian processes \cite{Rasmussen},
\begin{align}
\mathbf{y} = \mathbf{f} +\bm{\varepsilon}, \quad \mathbf{f}\sim\cN(\mathbf{0},\mathbf{K}), \quad \bm{\varepsilon}\sim\cN(\mathbf{0},\tau^2\mathbf{I}),\label{mod4}
\end{align}
where, in addition to previous notation, $\mathbf{f}=[f(\mathbf{x}_1),\ldots,f(\mathbf{x}_n)]^{\T}$ is assumed to come from a multivariate normal with mean $\mathbf{0}$ and covariance matrix $\mathbf{K} = \bm{\Psi}^{\T}\bm{\Psi}$ with each element $k_{ij} = k(\mathbf{x}_i,\mathbf{x}_j)$. Altogether, we refer to the family of models taking on this form as GP regression. The formulation of the weight space GP is similar to the linear mixed model (LMM) \cite{Lippert:2011aa,Zhou:2012aa} that is frequently used in genetics but with one key difference: the GP model utilizes a nonlinear covariance matrix $\mathbf{K}$ instead of the usual gram matrix $\mathbf{X}\mathbf{X}^{\T}/p$. From this perspective, an RKHS model can be viewed as an extension of the LMM for modeling nonlinear effects such as statistical interactions. Indeed, the GP model still presents the same modeling benefits as an LMM, such as controlling for structured random effects. For example, notice that the Gaussian covariance function can written as a product of three terms \cite{Cotter:2011aa}
\begin{align*}
\exp\left\{-\frac{1}{2\theta^2}\|\mathbf{x}-\mathbf{x}^\prime\|^2\right\} = \exp\left\{-\frac{1}{2\theta^2}\left(\|\bx\|^2+\|\bx^\prime\|^2\right)\right\}\exp\left\{-\frac{1}{2\theta^2}\bx^{\T}\bx^{\prime}\right\}.
\end{align*}
The last term includes (nonlinear transformed) elements of the LMM relatedness matrix that has been well known to effectively control for population stratification in genetic studies \cite{Kang:2010aa,Wu:2011,Yang:2014aa,Zhou:2014aa}. Because of these properties, RKHS-based models have become powerful tools for predictive problems in many research areas, and have been widely used for genomic selection in animal breeding programs \cite{Campos:2009aa,Campos:2010aa}. We replicate some of these sentiments via a small simulation study (see Supplementary Text and Table S1).

Lastly, we want to point out that (although not explicitly considered here) the formulation of the GP regression model in Equation \eq{mod4} can also be easily extended to accommodate other fixed effects (e.g.~age, sex, or genotype principal components) \cite{Campos:2009aa,Shi:2012aa}, as well as be adapted to account for interactions between variants and non-genetic risk factors \cite{Weissbrod:2016aa,Cuevas:2017aa}.

\paragraph{Note on Bandwidth Parameters.} In many cases, the covariance function is indexed by a bandwidth parameter $\theta$ (also known as a smoothing parameter or lengthscale) --- which we expansively write as $k_\theta(\bullet,\bullet)$. For example, the previously mentioned Gaussian kernel can be specified as $k_\theta(\mathbf{x},\mathbf{x}^\prime)=\exp\{-.5\|\mathbf{x}-\mathbf{x}^\prime\|^2/\theta^2\}$. Within a fully Bayesian model, this bandwidth parameter can be assigned a prior distribution and its posterior distribution may be inferred \cite{MJ:2011aa}. However for simplicity, we follow recent studies using the ``median heuristic'' and work with a fixed bandwidth that we choose as $\theta= \text{median}_{ij}\|\mathbf{x}_i-\mathbf{x}_j\|_2$ \cite{Chaudhuri:aa}.

\paragraph{Posterior Inference and Sampling.} We now briefly detail a simple Markov chain Monte Carlo (MCMC) sampling procedure for estimating the parameters in GP regression. Assume now that we have a completely specified hierarchical model
\begin{align*}
\mathbf{y} = \mathbf{f} + \bm{\varepsilon}, \quad \mathbf{f}\sim\mathcal{N}(\mathbf{0},\mathbf{K}), \quad \bm{\varepsilon}\sim\mathcal{N}(\mathbf{0},\tau^2\mathbf{I}), \quad \tau^2\sim\text{Scale-Inv-}\chi^2(a,b)
\end{align*}
where, in addition to previous notation, we further assume that the residual variance parameter $\tau^2$ follows a scaled-inverse chi-square distribution with degrees of freedom $a$ and scale $b$ as hyper-parameters. Given the conjugacy of this model specification, we may use a Gibbs sampler to estimate the joint posterior distribution $\mathcal{P}(\mathbf{f},\tau^2|\mathbf{y})$. This consists of iterating between the following two conditional densities:
\begin{enumerate}
\item $\mathbf{f}\cond\tau^2,\mathbf{y}\sim\mathcal{N}(\mathbf{m}^*,\mathbf{V}^*)$ where $\mathbf{m}^* = \bK(\bK+\tau^2\bI)^{-1}\by$ and $\mathbf{V}^* = \bK-\bK(\bK+\tau^2\bI)^{-1}\bK$;
\item $\tau^2\cond\mathbf{f},\mathbf{y}\sim\text{Scale-Inv-}\chi^2(a^*,b^*)$ where $a^* = a+n$ and $b^* = a^{*-1}[ab+(\mathbf{y}-\mathbf{f})^{\T}(\mathbf{y}-\mathbf{f})]$.
\end{enumerate}
Iterating the above procedure $T$ times results in a set of sampled draws from the target joint posterior distribution. Taking the mean over these draws yields posterior estimates for the model parameters (see Supplementary Material for a detailed algorithmic overview).

\subsection{Effect Size Analog for Nonparametric Methods}

A noteworthy downside to the GP regression model is the inability to find an effect size for causal variants. From a prediction and genomic selection perspective this loss is fine --- but from the perspective of finding genetic markers that give rise to this improved predictive performance (i.e.~association mapping), the interpretability of the model is lost. We now define the effect size analog for general nonparametric methods as a solution to this limitation \cite{Crawford:2017aa}. We first briefly outline the conventional wisdom for coefficients in linear regression. In linear models, a natural interpretation of a regression coefficient is the projection of the genotypes $\mathbf{X}$ onto the phenotypic vector $\mathbf{y}$, 
\begin{align}
\widehat{\bm{\beta}} = \mbox{Proj}({\mathbf X},\mathbf{y}),\label{eff}
\end{align}
with the choice of loss function, noise model, as well as prior distributions or regularization penalties, specifying the exact form of the projection. One standard projection operation is $\mbox{Proj}(\mathbf{X},\mathbf{y}) = \mathbf{X}^{\dagger} {\mathbf y}$, where $\mathbf{X}^{\dagger}$ is the Moore-Penrose generalized inverse. For Bayesian procedures, priors over the parameters $\bm{\beta}$ induce a distribution on the resulting projection procedure $\mbox{Proj}(\mathbf{X}, \mathbf{y})$ \cite{Liang:2008aa,Carvalho:2010aa}. 

The general definition for the effect size analog is based on the similar idea of projecting a nonlinear function onto the design matrix. Specifically, consider a nonlinear function evaluated on $n$-observed samples, such that $\E(\mathbf{y}\cond\mathbf{X}) = \mathbf{f}$. We formally define the \textit{effect size analog} as the result of projecting the genotypic matrix $\mathbf{X}$ onto the nonlinearly estimated function vector $\mathbf{f}$,  
\begin{align}
\widetilde{\bm{\beta}} = \mbox{Proj}({\mathbf X},{\mathbf f})\label{geff}.
\end{align}
This projection operation, and its practical calculation, effectively requires two sets of coefficients: (i) the theoretical coefficients $\mathbf{c}$ on the basis functions; and (ii) the coefficients that determine the effect size analog $\widetilde{\bm{\beta}}$. Following the formulation in Equation \eqref{geff}, we use Equations \eq{mod2} and \eq{mod4} to specify the joint projection of design matrix $\mathbf{X}$ onto the vector $\mathbf{f} = \bm{\Psi}^{\T} \mathbf{c}$ as the linear map,
\begin{align}
\widetilde{\bm{\beta}} = \mathbf{X}^{\dagger}\bm{\Psi}^{\T}\mathbf{c} = \mathbf{X}^{\dagger}\mathbf{f}.\label{eq36}
\end{align}
The argument for why the $p$-dimensional vector $\widetilde{\bm{\beta}}$ is an effect size analog for nonparametric regression models is that, on the $n$-observations, $\mathbf{f}\approx \mathbf{X}\widetilde{\bm{\beta}}$. In Supplementary Material, we re-derive previous results to formally show that the map from $\mathbf{f}$ to $\widetilde{\bm{\beta}}$ is injective modulo the null space of the genotypic matrix \cite{Crawford:2017aa}. This is similar to the classical linear regression case where two different coefficient vectors will result in the same estimated value if the difference between the vectors is in the null space of $\mathbf{X}$. Additionally, the only requirement for Equation \eqref{eq36} is a well-defined feature map $\psi(\bullet)$. This includes taking the Cholesky decomposition of the covariance matrix as a feature map, or even employing low-rank approximations such as the Nystrom approximation \cite{Drineas:2005}, random Fourier features \cite{Recht:2007aa}, or explicit Mercer expansions \cite{fasshauer2016kernel}. We should be clear that a variety of projection procedures (corresponding to various priors and loss functions) can be specified, and a systematic study elucidating which projections are efficient and robust is of interest for future research.

A key motivation for the effect size analog is to conduct nonlinear association mapping in the original genotype space, while also accounting for population structure and significant covarying relationships between variants. When a phenotype or trait is solely driven by additive effects, the projections \eq{eff} and \eq{geff} with the same genotypes $\mathbf{X}$ are equivalent, and the resulting effect size analog from Equation \eq{eq36} is the same as the OLS estimate derived by a standard linear model. Alternatively, it has been shown (via Taylor series expansions) that certain covariance functions enumerate nonlinear effects among observed markers \cite{Jiang:2015aa}. The Gaussian kernel, in particular, includes all higher-order interaction components, where the contribution of the terms decays polynomially with the order of nonlinearity \cite{Cotter:2011aa}. Therefore, when a given phenotype is driven by an arbitrary combination of additivity and interactions, a properly chosen nonlinear map $\psi(\bullet)$ will lead to an inversion in Equation \eq{eq36} that represents each $\widetilde{\beta}_j$ as a weighted sum of higher order interactions between marker $j$ and all other markers (see text in Supplementary Material).


\section{Genetic Association Mapping using Centrality Measures}\label{sec3}

The effect size analog serves as a nonlinear summary coefficient for each genetic variant in the original modeling space. However, since the explicit projection in Equation \eq{eq36} does not always guarantee a preserved mapping of sparse solutions \cite{Crawford:2017aa}, we cannot directly use standard Bayesian quantities such as posterior inclusion probabilities (PIPs) or Bayes factors (BFs) to rank markers in order of their significance. Indeed, there are many approaches to compute marginal association statistics based on corresponding effect size estimates \cite{Barbieri:2004aa,Stephens_Nature}, but many of these techniques rely on arbitrary thresholding. More importantly, they also fail to take advantage of significant underlying dependencies and covarying relationships between variants or sets of genomic loci. 

We now develop our main methodological innovation. We introduce an analogy to traditional Bayesian hypothesis testing for nonparametric regression methods: a \textit{post-hoc} approach for association mapping via a series of ``distributional centrality measures'' using Kullback-Leibler divergence (KLD) \cite{goutis1998model,Smith:2006aa,Woo:aa,Tan:2017aa,piironen2016projection,piironen2017comparison,Alaa:2017aa}. Our strategy will be to use the posterior samples of the effect size analogs to infer the relative covariance between genetic variants. This underlying correlation structure will then be systematically searched over to posit significant individual associations. We refer to this approach as computing the ``RelATive cEntrality'' of genetic markers, or RATE. 

\subsection{Kullback-Leibler Divergence} 

Typical questions in network studies simplify to the general issue of determining the ``centrality'' of nodes --- the potential importance of individual components in relation to the other nodes in the entire network. When network relationships are modeled via multivariate distributions, this can be explored in various statistical ways. Assume here that we have a collection of deterministically computed samples from the implied posterior distribution of the effect size analog $\widetilde{\bm{\beta}}$ (via the projection in Equation \eq{eq36}). One interpretable way to summarize (in a single measure) the influence/importance of the $j$-th variant in $\mathbf{x}_j$, on the rest of the variants in $\mathbf{X}_{-j}$, is via the computation of the KLD measuring the difference between $\mathcal{P}(\widetilde{\bm{\beta}}_{-j}\cond \widetilde{\beta}_j)$ and $\mathcal{P}(\widetilde{\bm{\beta}}_{-j})$. Specifically, this is defined by solving the following integral
\begin{align}
\text{KLD}(\widetilde{\beta}_j) = \int_{\widetilde{\bm{\beta}}_{-j}}\log\left(\frac{\mathcal{P}(\widetilde{\bm{\beta}}_{-j})}{\mathcal{P}(\widetilde{\bm{\beta}}_{-j}\cond \widetilde{\beta}_j)}\right) \mathcal{P}(\widetilde{\bm{\beta}}_{-j}) \diff\widetilde{\bm{\beta}}_{-j}, \quad \quad j = 1,\ldots,p,\label{KLD} 
\end{align}
where we use the shorthand $\text{KLD}(\widetilde{\beta}_j) = \text{KLD}(\mathcal{P}(\widetilde{\bm{\beta}}_{-j})\|\mathcal{P}(\widetilde{\bm{\beta}}_{-j}\cond \widetilde{\beta}_j))$. Here, the KLD is a non-negative quantity and, in this context, takes the value of zero if and only if $\mathcal{P}(\widetilde{\bm{\beta}}_{-j}\cond \widetilde{\beta}_j) = \mathcal{P}(\widetilde{\bm{\beta}}_{-j})$. Equivalently, this means that the KLD is zero if and only if the posterior distribution of $\widetilde{\bm{\beta}}_{-j}$ is independent of the effect $\widetilde{\beta}_j$. Therefore, the case for which $\text{KLD}(\widetilde{\beta}_j)=0$ may simply be interpreted as meaning that variant $j$ is not a key explanatory variable relative to others. Otherwise, for any given conditioning value $\widetilde{\beta}_j$, the divergence in Equation \eq{KLD} represents the information (i.e.~entropy) change induced on the distribution of $\widetilde{\bm{\beta}}_{-j}$ --- naturally varying as the conditioning value $\widetilde{\beta}_j$ varies. 

\paragraph{Closed Form Derivation under Approximate Normal Posteriors.} 
For our case study and immediate applications, we are interested in straightforward computation of KLD measures in order to address problems with increasingly large numbers of genotypes and possible interactions. For these purposes, and the rest of the paper, we therefore restrict attention to contexts in which we can assume an adequate normal approximation to the full joint posterior distribution of the $p$-dimensional effect size analog $\widetilde{\bm{\beta}}.$ Ongoing and future work is concerned with computational and numerical aspects of the more general context, while the methodological and applied advances enabled by our approach are well-highlighted under the normal posterior assumption. 

Thus, we take the posterior for $\widetilde{\bm{\beta}}$ as (approximately) multivariate normal with an empirical mean vector $\bm{\mu}$ and positive semi-definite covariance/precision matrices $\bm{\Sigma} = \bm{\Lambda}^{-1}$ estimated via simulation methods. Consider the association mapping case where we want to investigate the centrality or marginal importance of marker $j$. We may partition conformably as follows
\begin{align*}
\widetilde{\bm{\beta}} =  \beginmat{c}\widetilde{\beta}_j\\ \widetilde{\bm{\beta}}_{-j}\endmat, \quad \bm{\mu} = \beginmat{c}\mu_j\\ \bm{\mu}_{-j}\endmat, \quad \bm{\Sigma} = \beginmat{cc} \sigma_{j} & \bm{\sigma}^{\T}_{-j}\\ \bm{\sigma}_{-j} & \bm{\Sigma}_{-j}\endmat, \quad \bm{\Lambda} = \beginmat{cc} \lambda_{j} & \bm{\lambda}^{\T}_{-j}\\ \bm{\lambda}_{-j} & \bm{\Lambda}_{-j}\endmat,
\end{align*}
where $\widetilde{\beta}_j$, $\mu_j$, $\sigma_{j}$ and $\lambda_{j}$ are scalars; $\widetilde{\bm{\beta}}_{-j}$, $\bm{\mu}_{-j}$, $\bm{\sigma}_{-j}$, and $\bm{\lambda}_{-j}$ are $(p-1)$-dimensional vectors; and $\bm{\Sigma}_{-j}$ and $\bm{\Lambda}_{-j}$ are $(p-1)\times(p-1)$ positive definite, symmetric matrices. Under this partitioning, we know that the marginally $\widetilde{\bm{\beta}}_{-j}\sim\mathcal{N}(\bm{\mu}_{-j},\bm{\Sigma}_{-j})$. Furthermore, we also know that, when conditioned on the $j$-th variant, $\mathcal{P}(\widetilde{\bm{\beta}}_{-j}\cond\widetilde{\beta}_j)$ is a multivariate distribution with expectation and covariance
\begin{align*}
\mathbb{E}(\widetilde{\bm{\beta}}_{-j}\cond\widetilde{\beta}_j) = \bm{\mu}_{-j}+\bm{\theta}_{j}(\widetilde{\beta}_j-\mu_j), \quad \mathbb{V}(\widetilde{\bm{\beta}}_{-j}\cond\widetilde{\beta}_j) = \bm{\Lambda}_{-j}^{-1},
\end{align*}
where $\bm{\theta}_j = -\bm{\Lambda}_{-j}^{-1}\bm{\lambda}_{-j}$ is a $(p-1)$-dimensional vector. Inserting these probability density forms into Equation \eq{KLD}, with some algebraic rearrangement, yields the following
\begin{align}
\text{KLD}(\widetilde{\beta}_j) = \frac{1}{2}\left[-\log|\bm{\Sigma}_{-j}\bm{\Lambda}_{-j}| + \mathbb{E}(\mathbf{e}_{-j}^{\T}\bm{\Lambda}_{-j}\mathbf{e}_{-j})-2\mathbb{E}(\mathbf{e}_{-j}^{\T})\bm{\Lambda}_{-j}\btheta_j\text{e}_j -\mathbb{E}(\mathbf{e}_{-j}^{\T}\bm{\Sigma}^{-1}_{-j}\mathbf{e}_{-j})+\text{e}_j^2\bm{\theta}_j^{\T}\bm{\Lambda}_{-j}\bm{\theta}_j\right]\label{KLDcomplex}
\end{align}
where $\log|\bullet|$ represents the log determinant function of a matrix, $\mathbf{e}_{-j} = \widetilde{\bm{\beta}}_{-j}-\bm{\mu}_{-j}$ is a vector, $\text{e}_j = \widetilde{\beta}_j-\mu_j$ is a scalar, and the expectations are taken with respect to the marginal posterior distribution of $\widetilde{\bbeta}_{-j}$. Next, denote the following definition of an expectation of quadratic forms \cite{Mathai:1992aa},
\begin{align*}
\mathbb{E}(\mathbf{u}^{\T}\mathbf{Q}\mathbf{u}) = \mathbb{E}(\mathbf{u}^{\T})\mathbf{Q}\mathbb{E}(\mathbf{u})+\tr(\mathbb{V}(\mathbf{u})\mathbf{Q}),
\end{align*}
for any vector $\mathbf{u}$ and positive semi-definite covariance matrix $\mathbf{Q}$, where $\tr(\bullet)$ is the matrix trace function. Using this equality, the computation of the KLD in Equation \eq{KLDcomplex} simplifies to the following closed form
\begin{align}
\text{KLD}(\widetilde{\beta}_j) = \frac{1}{2}\left[-\log(|\bm{\Sigma}_{-j}\bm{\Lambda}_{-j}|) + \tr(\bm{\Sigma}_{-j}\bm{\Lambda}_{-j}) + 1 - p + \alpha_j(\widetilde{\beta}_j - \mu_j)^2\right]\label{KLsimple}, 
\end{align}
where $\alpha_j = \bm{\theta}_j^{\T}\bm{\Lambda}_{-j}\bm{\theta}_j = \bm{\lambda}^{\T}_{-j}\bm{\Lambda}_{-j}^{-1}\bm{\lambda}_{-j}$ and $\tr(\mathbf{I}) = p-1$. By symmetry in the notation for elements of sub-vectors and sub-matrices, it trivially follows that we may simply permute the order of the variables in $\widetilde{\bm{\beta}}$ and iteratively compute the KLD to measure the centrality of any variant $j$.

\subsection{Prioritization and Relative Significance}

In the nonlinear regression context, values $\widetilde{\beta}_j$ close to zero may be interpreted as ``null hypotheses'' with little to no relevance to the modeled outcome. Therefore, searching for the most central (i.e.~influential) genetic markers simply reduces to looking for the greatest KLD when setting each $\widetilde{\beta}_j = 0$. More contextually specific questions arise when deciding if a given centrality measure is significant. Indeed, in practice, a threshold may be chosen in order to determine if any given KLD represents a significant shift in entropy. Previous studies have done this through $k$-fold permutation to find an effective genome-wide threshold \cite{Woo:aa}. This approach can be costly for datasets with many markers.

A more computationally efficient option for determining a natural ranked cutoff is to explore the relevance of variables recursively, and judge their significance via a scaled version of the KLD. We call this ``RelATive cEntrality'' or RATE,
\begin{align}
\text{RATE}(\widetilde{\beta}_j) = \text{KLD}(\widetilde{\beta}_j)/\sum\text{KLD}(\widetilde{\beta}_\ell), \quad \sum\text{RATE}(\widetilde{\beta}_j)=1.\label{RATE}
\end{align}
Here, the RATE measure is bounded within the range $[0,1]$, with the natural interpretation of measuring a variable's relative importance. Suppose that $j$ identifies the genetic marker with the largest RATE value. Conditioning on a reduced margin, and then repeating the computation outlined in Equations \eq{KLsimple} and \eq{RATE}, will identify the relatively second most explanatory marker. We can repeat this procedure until each of the remaining variants appear to be equal in their relative importance. This would indicate that all significant variants had been identified, and all that remain are variants for which their influences on the posterior distribution are indistinguishable. This recursive process can be simplified to defining an initial set of candidate associated markers with first order centrality measures satisfying 
\begin{align*}
\left\{j: \text{RATE}(\widetilde{\beta}_j)> 1/p\right\}.
\end{align*}
The value $1/p$ represents the null assumption that there is relatively equal importance across all variants; and hence, there are no central nodes that exist within the posterior distribution. We may quantify this behavior by checking the entropic difference between a uniform distribution and the observed RATE measures. Namely,
\begin{align}
\Delta = \log(p) - H, \quad H = -\sum\text{RATE}(\widetilde{\beta}_j)\log(\text{RATE}(\widetilde{\beta}_j)),\label{Delta}
\end{align} 
where $H$ represents the intrinsic entropy of the relative centrality measures, and the case of no significantly associated markers yields an entropy of $\log(p)$. One way to calibrate $\Delta$ is linked to effective sample size (ESS) measures from importance sampling \cite{GruberWest2016BA,GruberWest2017ECOSTA}. In a very different applied context, authors have exploited the use of an approximate ESS measure defined by 
\begin{align}
\text{ESS} = 1/(1+\Delta)\times100\%.\label{ESS}
\end{align}
This ESS measure is a calibration metric that provides a notion of ``loss in uniformity''. For example, 50\% loss in terms of (1-ESS) translates to a larger $\Delta$ value of 1. This equates to the presence of at least one variant that is significantly associated with the observed phenotypic trait. On the other hand, a minor 5\% loss corresponds to a more uniform case with $\Delta$ value of about 0.05. Again, this latter scenario would occur when there are hardly any influential markers within the data. 

For any given set of significant variables, according to their estimated RATE measure, further analyses may be carried out involving the relative costs of false positives and negatives to make an explicitly reasoned decision about which specific variants to pursue \cite{Stephens_Nature}. Unless stated otherwise, the results we present throughout the rest of the paper will be based on using RATE. We explore the power of this alternative approach for association mapping in Section \ref{sec4}.

\subsection{Relationship to Graphical Models and Precision Analysis}

In conventional statistics, the proposed variable selection procedure is very much related to precision analysis. It follows that the rate of change for the KLD (i.e.~the first derivative of Equation \eq{KLsimple} with respect to a given effect size analog), is found via the term $\alpha_j = \bm{\lambda}^{\T}_{-j}\bm{\Lambda}_{-j}^{-1}\bm{\lambda}_{-j}$. This means that the closed form computation of the KLD is directly impacted by the deviations between the approximation of a given predictor's posterior mean and the assumption that its true effect is zero. Therefore, $\alpha_j$ characterizes the implied linear rate of change of information when the effect of any predictor is absent --- thus, providing a natural (non-negative) numerical summary of the role of $\widetilde{\beta}_j$ in the multivariate distribution. In terms of weightings from the precision matrix, we see the following equivalent representation for the rate of change of the KLD,
\begin{align*}
\alpha_j = \sum_{k\ne j}\sum_{\ell\ne j}c_{k\ell}\lambda_{jk}\lambda_{j\ell},
\end{align*}
where $c_{k\ell}$ is the corresponding $k$-$\ell$-th element of the matrix $\bm{\Lambda}_{-j}^{-1}$. As derived in the previous subsection, we may alternatively denote $\alpha_j = \bm{\theta}_{-j}^{\T}\bm{\Lambda}_{-j}\bm{\theta}_{-j}$, where again $\bm{\theta}_{-j} = -\bm{\Lambda}_{-j}^{-1}\bm{\lambda}_{-j}$ is a $(p-1)$-dimensional vector and $\bm{\Lambda}_{-j}$ is the precision matrix of the conditional distribution $\mathcal{P}(\widetilde{\bm{\beta}}_{-j}\cond \widetilde{\beta}_j)$. These representations help show that, in the context of normal statistical regression, $\alpha_j$ computes the ``variance explained'' (i.e.~the fitted sum-of-squares) by each covariate $j$.

The idea of variable selection via entropic shifts also has a key connection to graphical models. Often the goal of graphical models is to investigate if the precision matrix has some off-diagonal series corresponding to an underlying conditional independence structure between predictor \cite{Carvalho:2007aa}. RATE --- a relative distributional centrality measure that assesses importance (or influence) of each variable on the network of relationships reflected in the graph --- is greatly affected by the graphical structure resulting from the implied zeros in $\bm{\Lambda}$. A missing edge between two predictors $j$ and $\ell$ means that $\lambda_{j\ell} = 0$; hence, limiting the contribution of node $\ell$ to the overall ``network impact factor'' of $\alpha_j$. From the sum defining $\alpha_j$ above, we see that a term related to variables $k$ and $\ell$ is non-zero only when both $\lambda_{jk}$ and $\lambda_{j\ell}$ are non-zero. Therefore, the $k$-$\ell$-th summation term is non-zero only for pairs of predictors that are direct neighbors of $j$ in an undirected graph.

\subsection{Software Implementation}

Software for computing the RATE measure is carried out in R code, which is freely available at \url{https://github.com/lorinanthony/RATE}. Detailed derivations of the algorithm, which utilizes low-rank matrix factorizations for a more practical implementation, are derived in Supplementary Material. 



\section{Results}\label{sec4}

We now illustrate the utility of using centrality measures for genetic association mapping through extensive simulation studies and real data analyses. The motivation for each set of examples is to better understand the performance and behavior of RATE under different types of genetic architectures. First, we use a small simulation study to help the reader build a stronger intuition about how RATE prioritizes influential variables in a dataset. It is during this demonstration where we also explore what happens to the concept of ``centrality'' and ``uniformity'' when the effects of all known significant markers are assumed to be absent from the model. Next, we use more realistic simulations to assess the mapping power of our approach in genetic based applications. Here, the goal is to show that RATE performs association mapping as well as the most commonly used Bayesian and regularization modeling techniques. Finally, we assess the potential of the our approach in two real datasets. The first is an \textit{Arabidopsis thaliana} QTL mapping study consisting of six different metabolic traits from an F6 Bay-0 $\times$ Shahdara recombinant inbred lines (RILs) population. The second is a genome-wide association study (GWAS) in a heterogeneous stock of mice from the Wellcome Trust Centre for Human Genetics.

\subsection{Simulation Studies}

For all synthetic demonstrations and assessments, we consider a simulation design that is often used to explore the utility of statistical methods across different genetic architectures underlying complex phenotypic traits \cite{Crawford:2017ab,Zeng:2017aa,Crawford:2018aa}. First, we assume that all of the observed genetic effects explain a fixed proportion of the total phenotypic variance. This proportion is referred to as the ``broad-sense heritability'' of the trait, which we denote as $\mbox{H}^2$. From the more conventional statistics perspective, the parameter $\mbox{H}^2$ can alternatively be described as a factor controlling the signal-to-noise ratio in the simulations. Next, we use a genotypic matrix $\mathbf{X}$ with $n$ samples and $p$ single nucleotide polymorphisms (SNPs) to generate synthetic real-valued phenotypes that mirror genetic architectures affected by a combination of linear (additive) and interaction (epistatic) effects. 

We randomly choose a select subset of $j^*$ ``causal'' (or truly associated) SNPs as the determining factors of the data generating process. The linear effect sizes for all $j^*$ associated genetic variants are assumed to come from a standard normal distribution: $\beta_{j^*}\sim \mathcal{N}(0,1)$. When applicable, we also create a separate matrix $\mathbf{W}$ which holds all pairwise interactions between the causal SNPs. These corresponding interaction effect sizes are drawn as $\bm{\gamma}\sim \mathcal{N}(\bm{0},\mathbf{I})$. We scale both the additive and interaction effects so that collectively they explain a fixed proportion of $\mbox{H}^2$. Namely, the additive effects make up $\rho$\%, while the pairwise interactions make up the remaining $(1-\rho)$\%. Alternatively, the proportion of the heritability explained by additivity is said to be $\V(\mathbf{X}\bm{\beta})= \rho \mbox{H}^2$, while the proportion detailed by nonlinearity is given as $\V(\mathbf{W}\bm{\gamma})=(1-\rho)\mbox{H}^2$. We consider two choices for the parameter $\rho = \{0.5,1\}$. Intuitively, $\rho = 1$ represents the limiting case where the variation of a trait is driven by solely additive effects. For $\rho = 0.5$, the additive and interaction effects are assumed to equally contribute to the total phenotypic variance. Once we obtain the final effect sizes for all causal variants, we draw normally distributed random errors as $\bm{\varepsilon}\sim\mathcal{N}(\bm{0},\mathbf{I})$ to make up the remaining $(1-\mbox{H}^2)\%$ of the total $\V(\mathbf{y})$. Finally, continuous phenotypes are then created by summing over all observed effects using two simulation models: 
\begin{enumerate}[(i)]
\item Standard Model: $\mathbf{y} = \mathbf{X}\bm{\beta}+\mathbf{W}\bm{\gamma} +\bm{\varepsilon}$
\item Population Stratification Model: $\mathbf{y} = \mathbf{Z}\bomega + \mathbf{X}\bm{\beta}+\mathbf{W}\bm{\gamma} +\bm{\varepsilon}$ 
\end{enumerate}
where $\bZ$ contains covariates representing additional population structure, and $\bomega$ are the corresponding fixed effects which are also assumed to follow a standard multivariate normal distribution. Alternatively, one can think of the combined effect of $\mathbf{Z}\bomega$ as structured noise. To this end, simulations under model (ii) will make the appropriate assumption that $\V(\bZ\bomega)+\V(\bvarepsilon) = (1-\mbox{H}^2)$. For any simulations conducted under model (ii), genotype PCs are not included in any of the model fitting procedures, and no other preprocessing normalizations were carried out to account for the added population structure.

It is helpful to point out here that the main purpose of the following simulations is to demonstrate the utility of RATE in providing an explicit ranking of variable importance --- so as to uncover the implicit ranking assigned by nonparametric regression methods. Our simulation comparisons are thus targeted to illustrate how RATE can be used in this task, and how its overall variable selection performance differs from standard parametric mapping procedures in different scenarios.

\subsubsection{Proof of Concept Simulations: Demonstrating Centrality} 

In this subsection, we show how distributional centrality measures may be used and interpreted when prioritizing genetic markers in an association mapping study. Our main concern is to familiarize the reader with the behavior and concepts underlying RATE. To do this, we make use of $n=2000$ synthetic genotypes that are independently generated to have $p=25$ single nucleotide polymorphisms (SNPs) with allele frequencies randomly sampled from a uniform distribution over values ranging from $[0.05, 0.5]$. The resulting $n\times p$ simulated genotype matrix $\bX$ is then used to create continuous phenotypes using the standard generative model (i). Here, we assume that only the last three variants $j^* = \{23,24,25\}$ are causal, and that their combined genetic effects make up a moderate $\mbox{H}^2 = 60\%$ of the total phenotypic variation. We then examine the full two cases for the parameter $\rho = \{0.5,1\}$. As a brief reminder, $\rho$ represents the proportion of broad-sense heritability that is contributed by additivity versus interaction effects. Indeed, these simulation assumptions are not realistic in terms of the qualities observed in real data applications; however, we stress that this section merely serves as a simple demonstration of ``centrality'' and ``uniformity.'' The small number of variants allows us to clearly illustrate and visualize these proofs of concepts.

Throughout the rest of this subsection, we detail the behavior RATE in the simple linear case with $\rho=1$. Similar results for $\rho = 0.5$ can be found in Supplementary Material. For each simulation, we fit a standard GP regression model under a zero mean prior and a Gaussian covariance function using a Gibbs sampler with 10,000 MCMC iterations and hyper-parameters set to $a = 5$ and $b = 2/5$. During each iterate, a corresponding nonlinear projection is computed as in Equation \eq{geff}. This results in an approximation of the implied posterior distribution for the effect size analog. With these conditional draws, we calculate the distribution's empirical posterior mean, covariance, and precision. Next, we use the closed form solutions in Equations \eq{KLsimple} and \eq{RATE} to derive a RATE measure for each genetic marker.

\begin{figure}[H]
\centering
\subfigure[First Order Centrality]{
   \includegraphics[width = 0.475\textwidth]{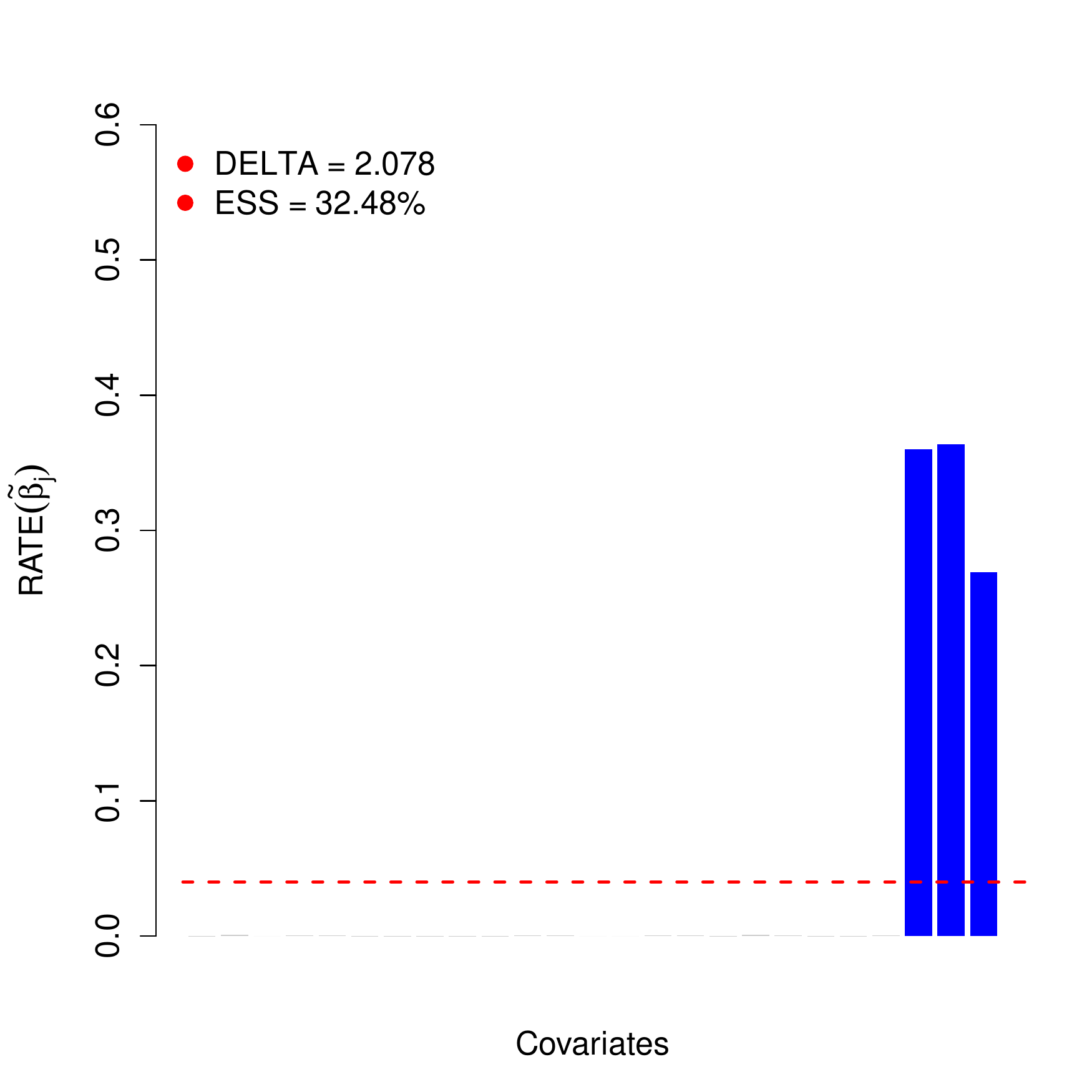}
   \label{Fig1A}
 }
 \subfigure[Second Order Centrality]{
   \includegraphics[width = 0.475\textwidth]{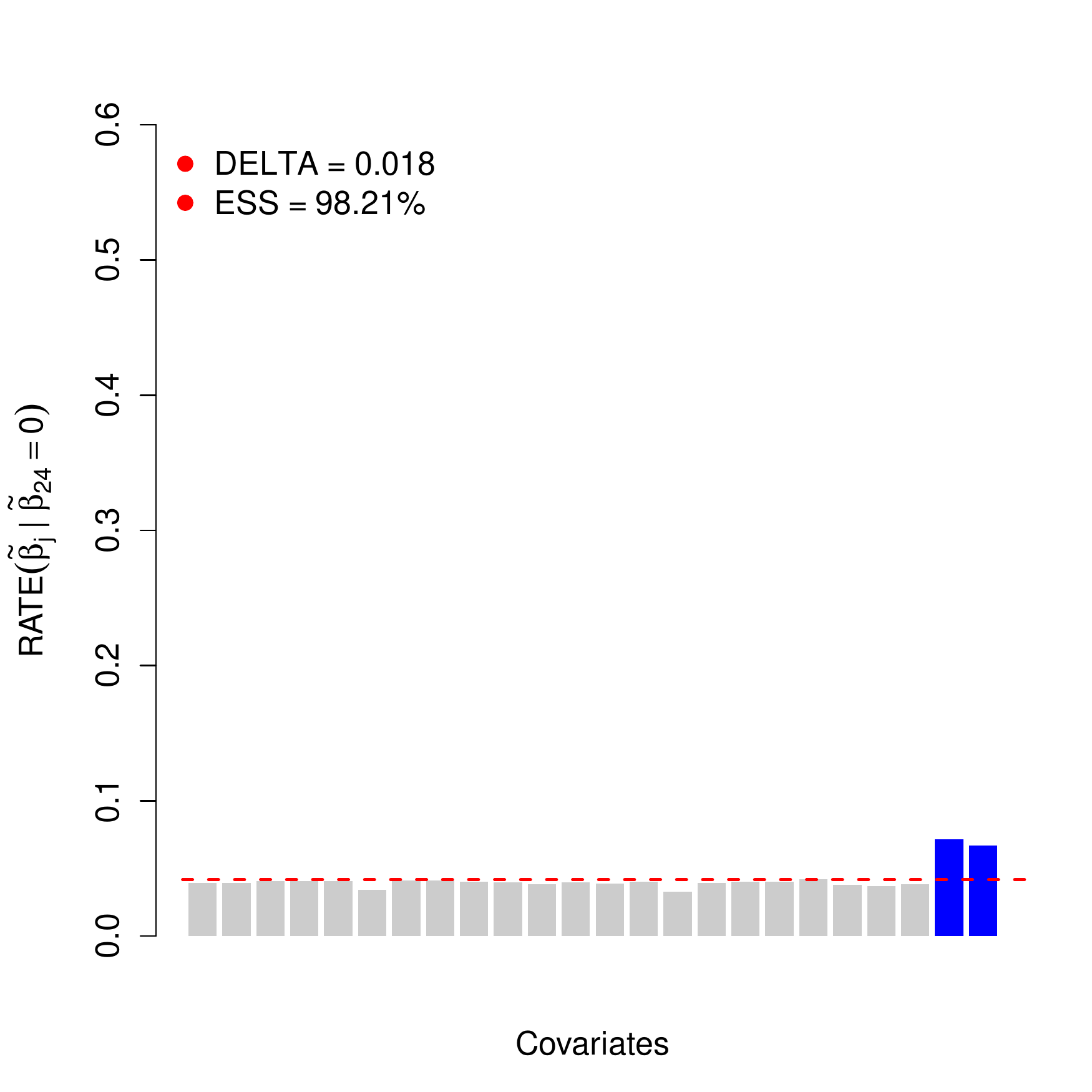}
   \label{Fig1B}
 }
 \subfigure[Third Order Centrality]{
   \includegraphics[width = 0.475\textwidth]{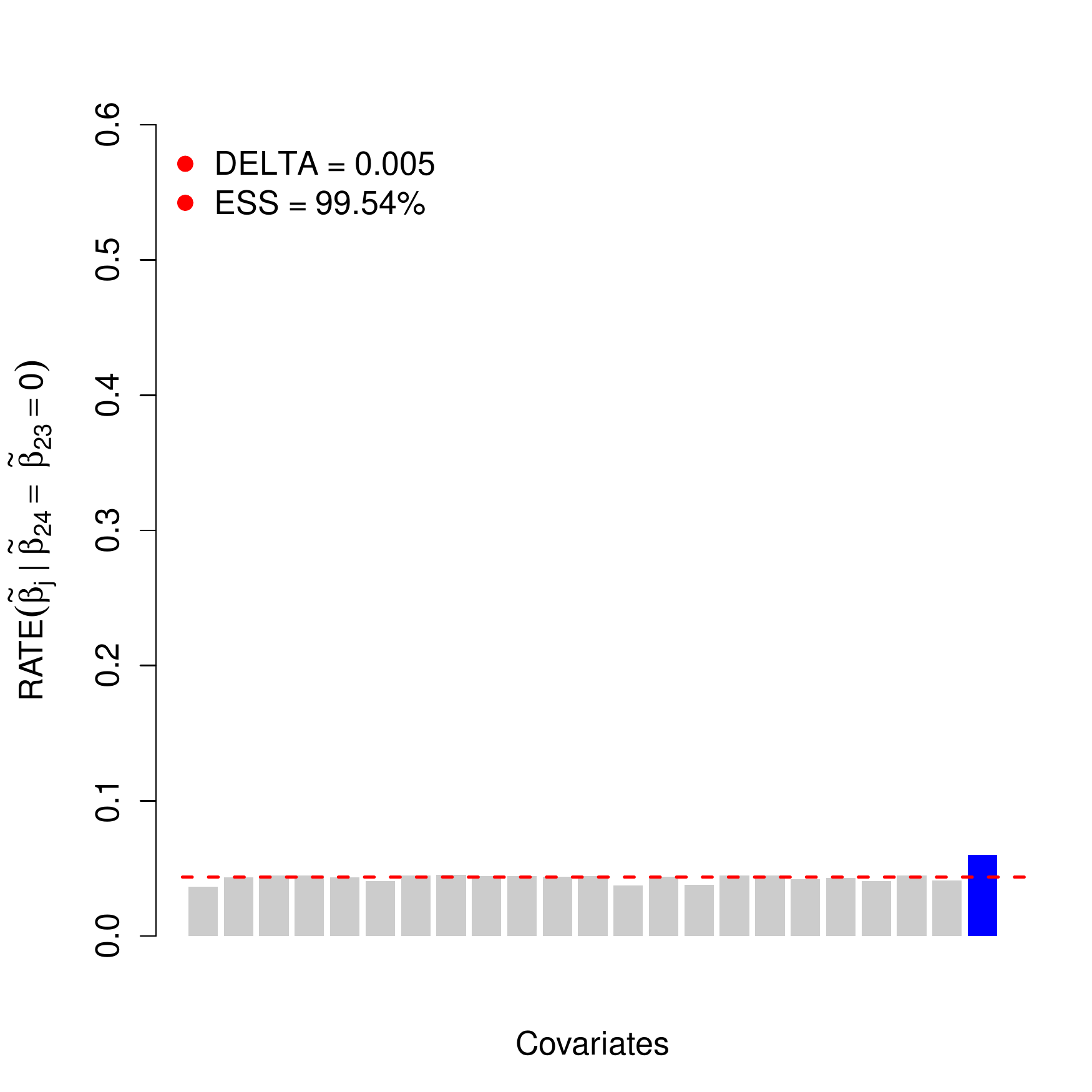}
   \label{Fig1C}
 }
 \subfigure[Fourth Order Centrality]{
   \includegraphics[width = 0.475\textwidth]{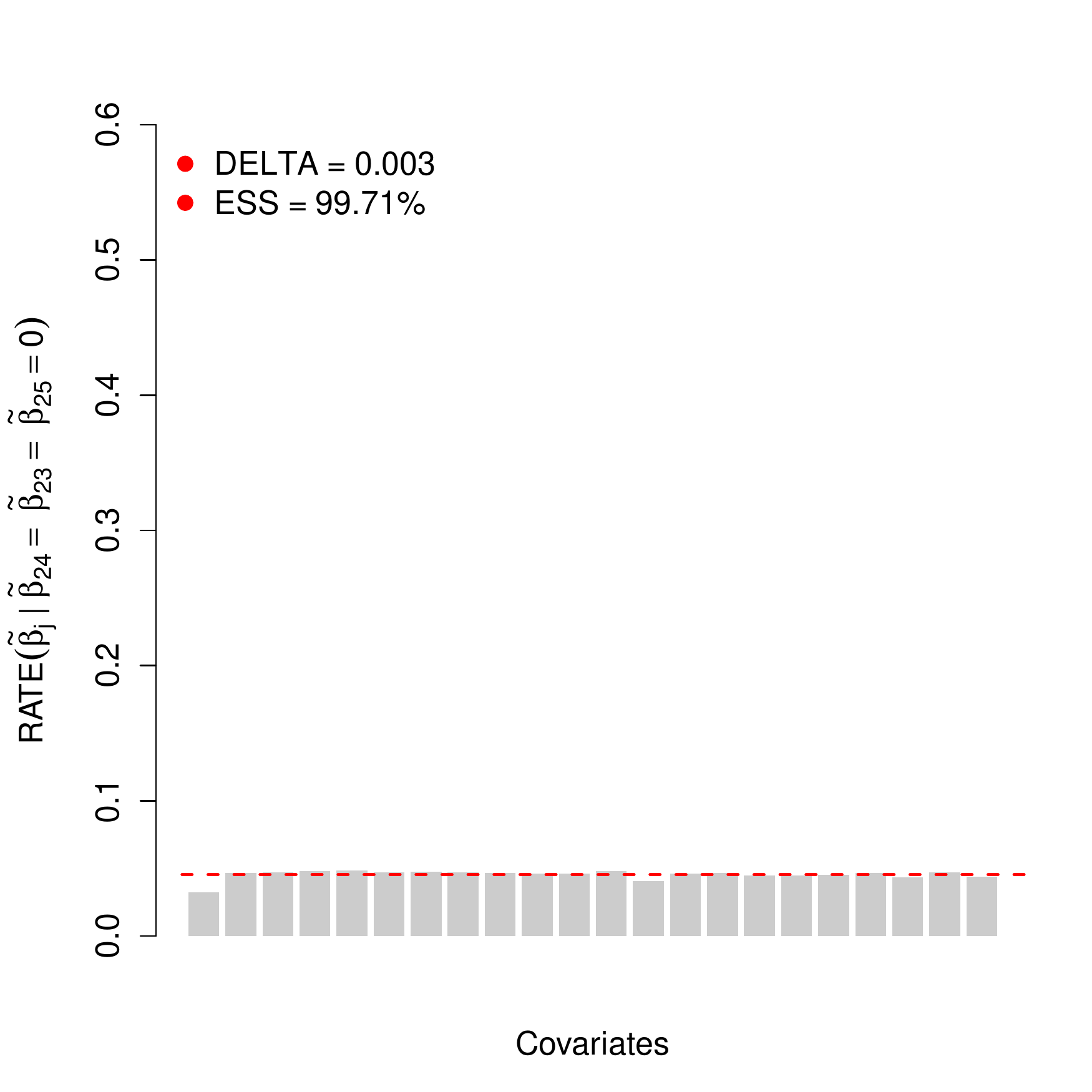}
   \label{Fig1D}
 }
\caption{\textbf{Orders of distributional centrality via RATE measures.} These are simple proof of concept simulations with broad-sense heritability level $\mbox{H}^2 = 0.6$ and $\rho = 1$. Here, $(1-\rho)$ is used to determine the proportion of signal that is contributed by interaction effects. Data are simulated such that the effects of only the last three genetic variants $j^* = \{23,24,25\}$ (blue) are nonzero. The dashed line is drawn at the level of relative equivalence (i.e.~$1/p$). Figure (a) shows the first order centrality across all markers; (b)-(d) show results when the most significantly associated variants are iteratively nullified. Uniformity check values are also reported: (i) the entropic difference $\Delta$, and (ii) the corresponding empirical effective sample size (ESS) estimates.}
\label{Fig1}
\end{figure}

Figure \ref{Fig1A} depicts an illustration of first order centrality across the 25 variants. Here, the three known causal SNPs are colored in blue. As a reference, we also display a red dashed line that is drawn at the level of relative equivalence (i.e.~$1/p$). This represents the value for which all variants are approximately uniform in their centrality or significance. To put this into better context, we provide uniformity checks: (i) the entropic difference $\Delta$ according to Equation \eq{Delta}, and (ii) the corresponding empirical ESS estimate as computed in Equation \eq{ESS}. In this first panel figure, we see that RATE accurately determines variants \#23-25 as being the most central to the posterior distribution. 

To demonstrate what it conceptually means to be central to a distribution, we next consider a series of follow-up analyses. Here, we iteratively assume that the genetic effect of the most significantly associated SNP has been nullified from the dataset. We then condition on a reduced margin for the posterior distribution and recompute the RATE measures. The key takeaway is that, without the effect of the data's most influential SNPs, the relative importance of the remaining variants will continue to increase until each of them are approximately equal in weight --- hence, resembling a uniform distribution. Consider the ongoing example, and assume that we nullify the effect of variant \#24. After recomputing $\text{RATE}(\widetilde{\beta}_j\cond\widetilde{\beta}_{24}=0)$ for every $j\ne24$-th variant, we see that while markers \#25 and \#23 are still the most significant according to their second order centrality, the importance levels of the other markers have shifted closer to becoming relatively equivalent (see Figure \ref{Fig1B}). This shift continues when the effects of the remaining causal variants are also removed successively (see Figures \ref{Fig1C} and \ref{Fig1D}, respectively). Also notice during this transition, $\Delta\rightarrow0$ and $\mbox{ESS}\rightarrow100\%$. This same trend happens both in the presence of interaction effects (Figure S2), as well as when the causal variants are in nearly perfect collinearity (or ``linkage disequilibrium'', LD) with non-causal markers (Figures S3 and S4). In the latter case, we force variants \#23-25 to have a correlation coefficient $R=0.9$ with variants \#1-3.

\begin{figure}[H]
\centering
\subfigure[First Order Centrality]{
   \includegraphics[width = 0.48\textwidth]{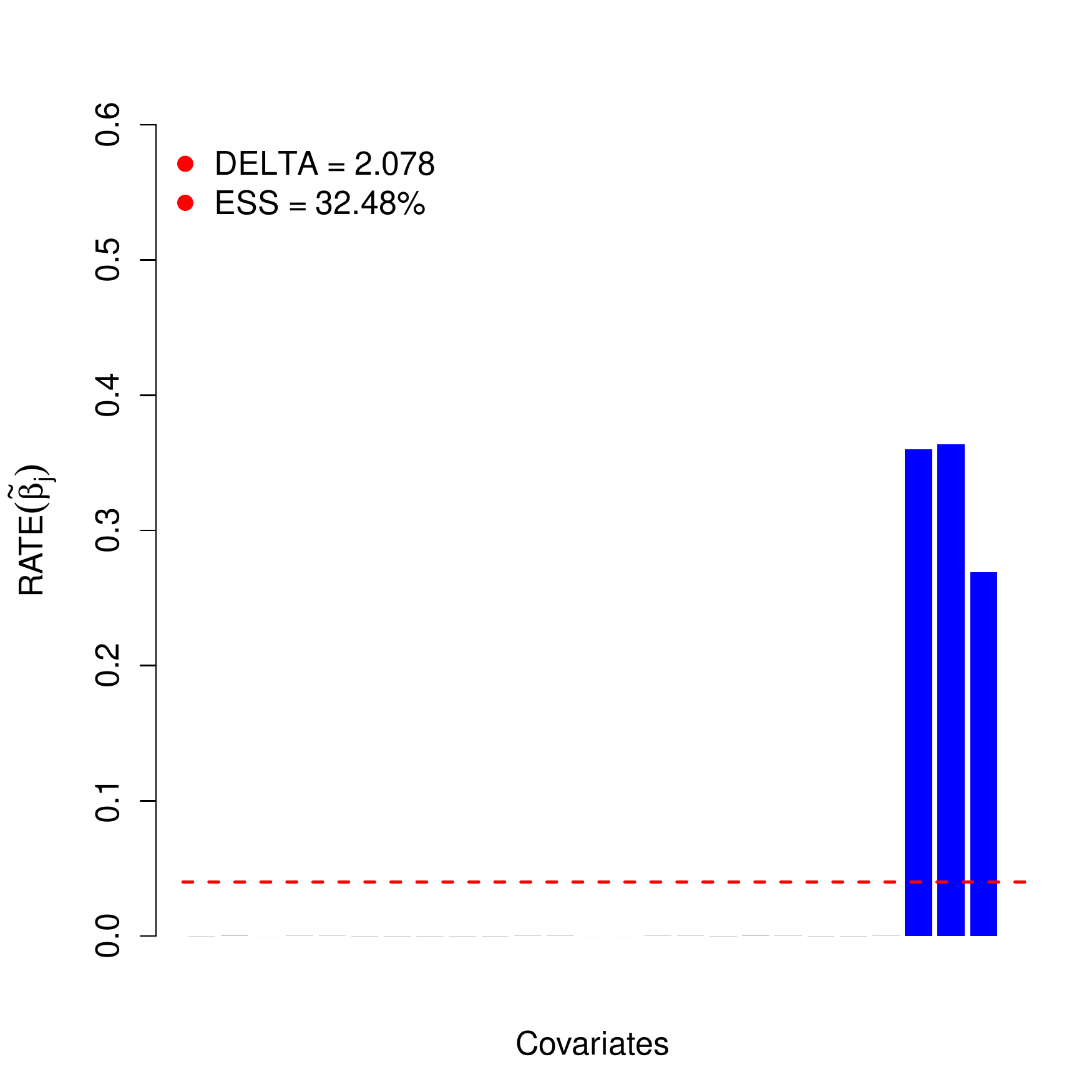}
   \label{Fig2A}
 }
 \subfigure[Second Order Centrality]{
   \includegraphics[width = 0.48\textwidth]{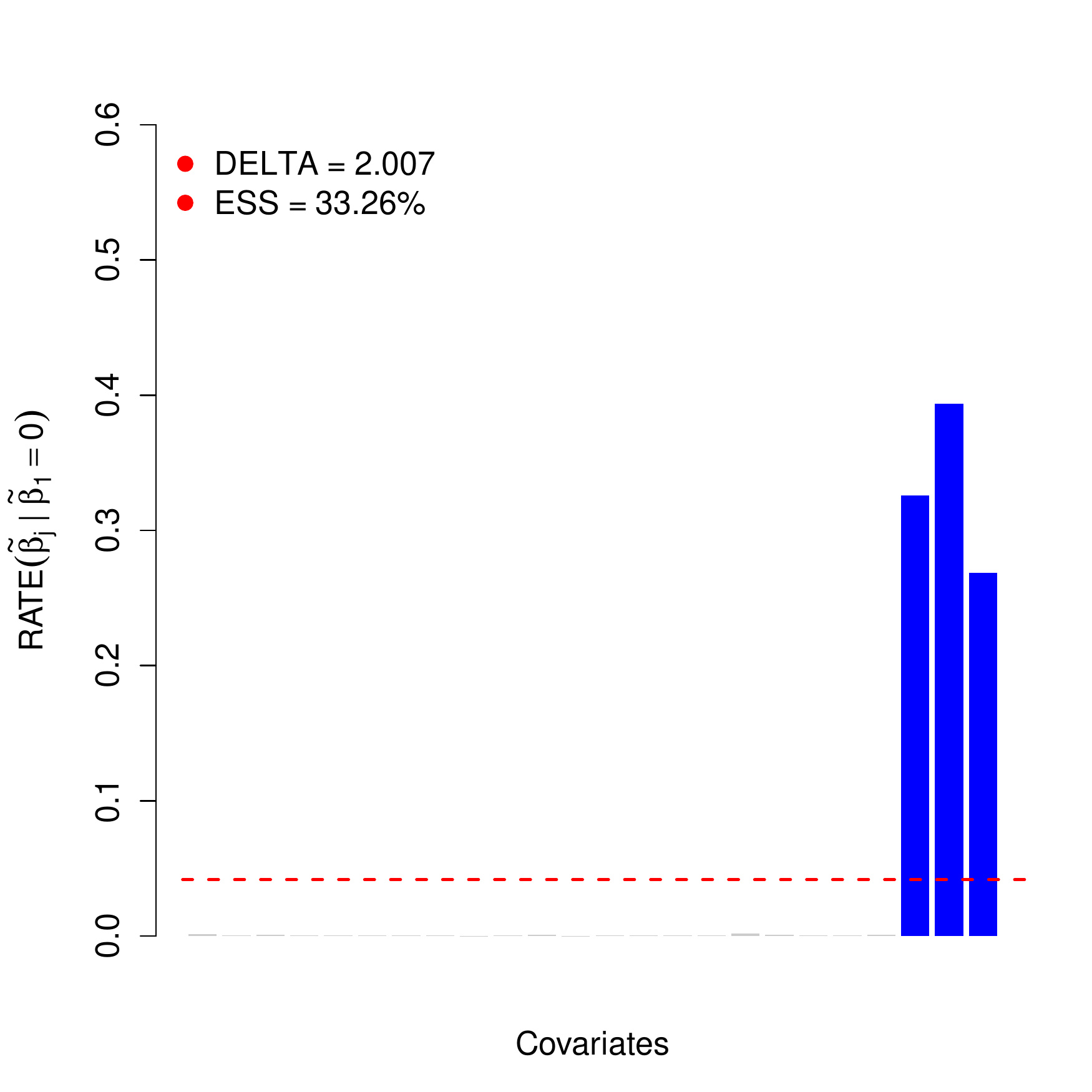}
   \label{Fig2B}
 }
 \subfigure[Third Order Centrality]{
   \includegraphics[width = 0.48\textwidth]{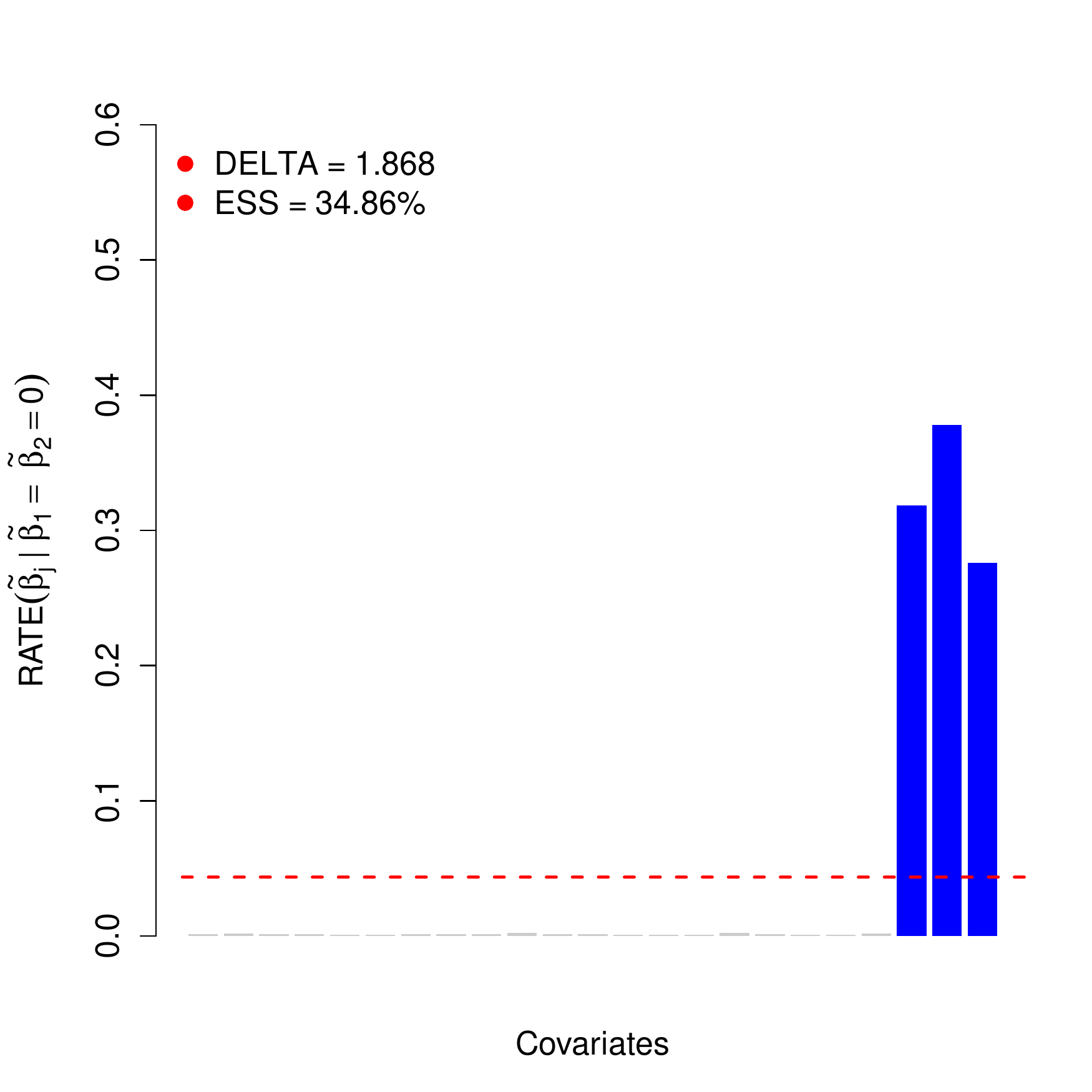}
   \label{Fig2C}
 }
 \subfigure[Fourth Order Centrality]{
   \includegraphics[width = 0.48\textwidth]{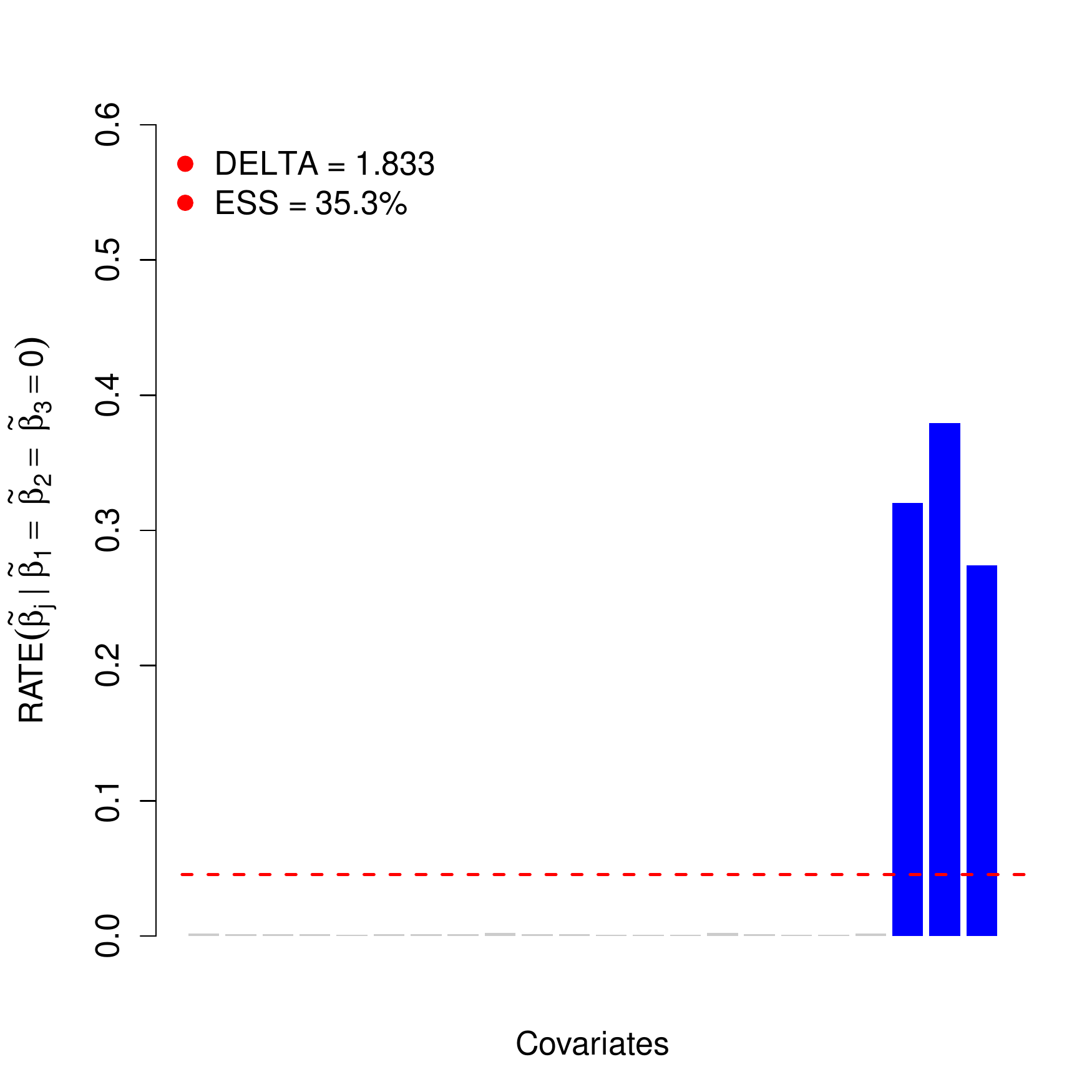}
   \label{Fig2D}
 }
\caption{\textbf{Orders of distributional centrality via RATE measures when non-associated variants are deemed significant.} These are simple proof of concept simulations with broad-sense heritability level $\mbox{H}^2 = 0.6$ and $\rho = 1$. Here, $(1-\rho)$ is used to determine the proportion of signal that is contributed by interaction effects. Data are simulated such that the effects of only the last three genetic variants $j^* = \{23,24,25\}$ (blue) are nonzero. The dashed line is drawn at the level of relative equivalence (i.e.~$1/p$). Figure (a) shows the first order centrality across all markers; (b)-(d) show the results when nonsignificant markers \#1-3 are iteratively nullified. Uniformity check values are also reported: (i) the entropic difference $\Delta$, and (ii) the corresponding empirical effective sample size (ESS) estimates.}
\label{Fig2}
\end{figure}

It is also important to demonstrate happens to the proposed centrality measures if one mistakenly removes the effect of a genetic marker that is not central to explaining the observed phenotypic variation. Reconsider the ongoing example where, instead of iteratively removing the effect of the most central variant, we simply nullify the effect of markers \#1-3 which we know to be nonsignificant (see Figure \ref{Fig2A}). Figures \ref{Fig2B}-\ref{Fig2D} (and Figure S5) illustrate that the three true causal variants (i.e.~markers \#23-25) are continuously identified as the most associated or central to the overall posterior distribution. Noticeably, with each passing removal of a non-central variant, the degree to which the RATE measures begin to look uniform has slowed substantially.

As a final demonstration, we show what happens when the null assumptions of relative centrality are met. Recall that, under the null hypothesis, RATE assumes that every variant equally contributes to the broad-sense heritability of a trait --- that is, no one SNP is more important or more central than the others. To illustrate this, we generate synthetic phenotypes such that the effect sizes of all twenty-five SNPs in the data are set to 1. Figure S6 shows results from four different datasets. The key takeaway here is that, in these cases, RATE produces much more uniformly distributed first order centrality measures as indicated by the entropic statistics $\Delta$ and $\mbox{ESS}$. For completeness, in Figure S7, we also show what happens to the raw and unscaled KLDs when phenotypes have been permuted. 

\subsubsection{Power Assessment and Method Comparisons}

We now assess the power of RATE and its ability to effectively prioritize truly associated variants under different genetic architectures. To do this, we now consider simulations that mirror more realistic genetic applications. Here, we utilize real genotypes from chromosome 22 of the control samples in the Wellcome Trust Case Control Consortium (WTCCC) 1 study \cite{WTCCC} (\url{http://www.wtccc.org. uk/}) to generate continuous phenotypes (see Supplementary Material for details). Exclusively considering this group of individuals and SNPs leaves us with an initial dataset consisting of $n =$ 2,938 samples and $p =$ 5,747 markers. During each simulation run, we randomly choose $j^* = 30$ SNPs which we classify into the two distinct causal groups: (1) a small set of 5 variants, and (2) a larger set of 25 variants. All causal markers have additive effects and, when applicable, the group 1 causal SNPs interact with group 2 causal SNPs, but never with each other (the same rule applies to the second group). We will consider three simulation scenarios. Scenario I involves phenotypes generated by standard model (i); while scenarios II and III consider model (ii) where we introduce population stratification effects by allowing the top 5 and 10 genotype principal components (PCs) $\bZ$ to make up 30\% of the overall variation in the simulated traits, respectively. Within these three scenarios, we set the broad-sense heritability to be $\mbox{H}^2= 0.3$ and consider two choices for the parameter $\rho = \{0.5,1\}$.

We compare the GP regression model and our proposed distributional centrality measures to a list of standard Bayesian and regularization modeling techniques. Specifically, these methods include: (a) a genome scan with a single-SNP univariate linear model that is typically used in GWAS applications (SCANONE) \cite{Yandell:2007aa} (b) L1-regularized lasso regression; (c) the combined regularization utilized by the elastic net \cite{Waldmann:2013aa}; and (d) a commonly used spike and slab prior model, also commonly known as Bayesian variable selection regression \cite{Guan:2011aa}, which places a prior distribution on each SNP as a mixture of a point mass at zero and a diffuse normal centered around zero. For each Bayesian method, we run a Gibbs sampler for 10,000 MCMC iterations. Regularization approaches were fit by first learning tuning parameter values via 10-fold cross validation.

\begin{figure}[H]
\centering
\subfigure[Scenario I ($\rho = 0.5$)]{
   \includegraphics[width = 0.48\textwidth]{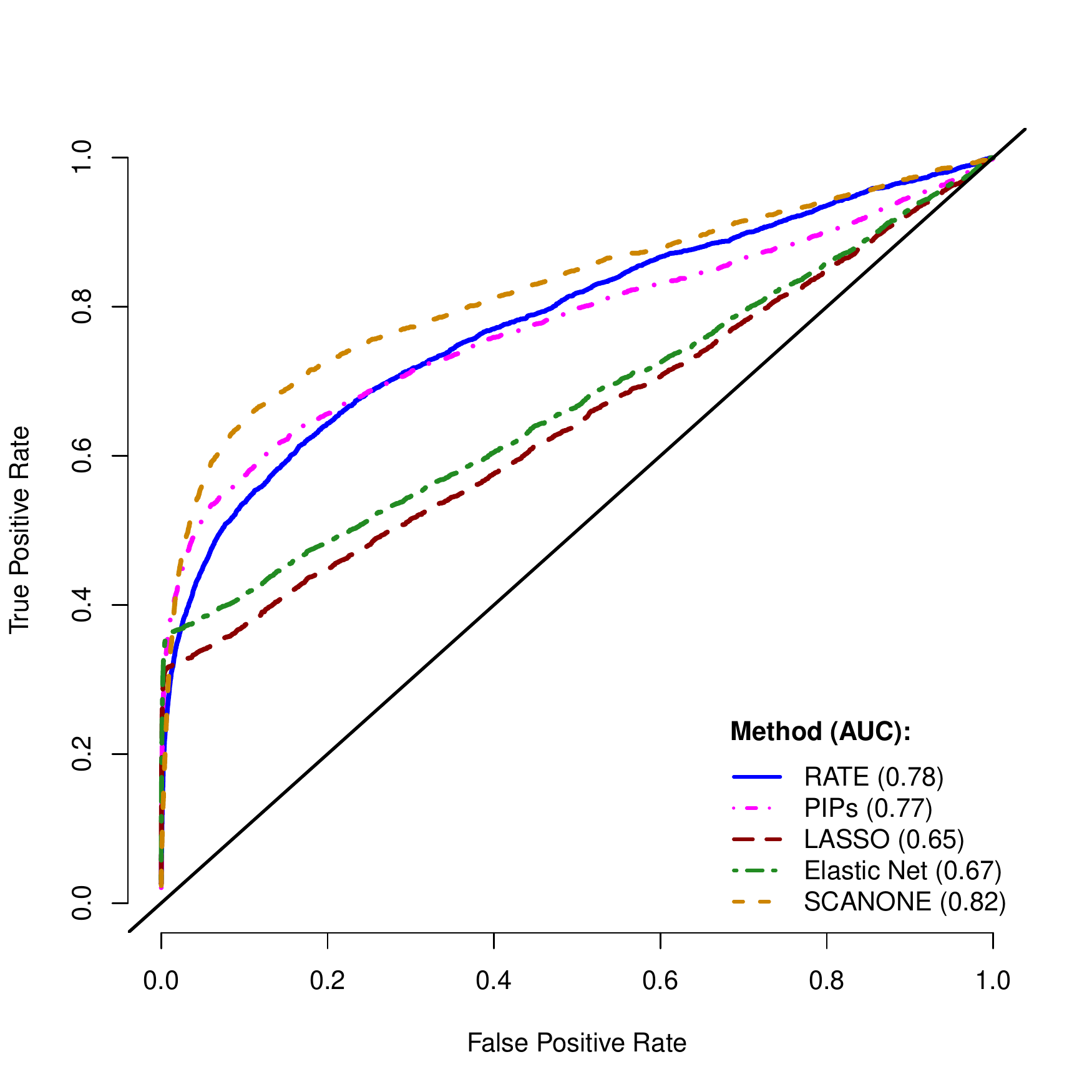}
   \label{Fig3A}
 }
 \subfigure[Scenario I ($\rho = 1$)]{
   \includegraphics[width = 0.48\textwidth]{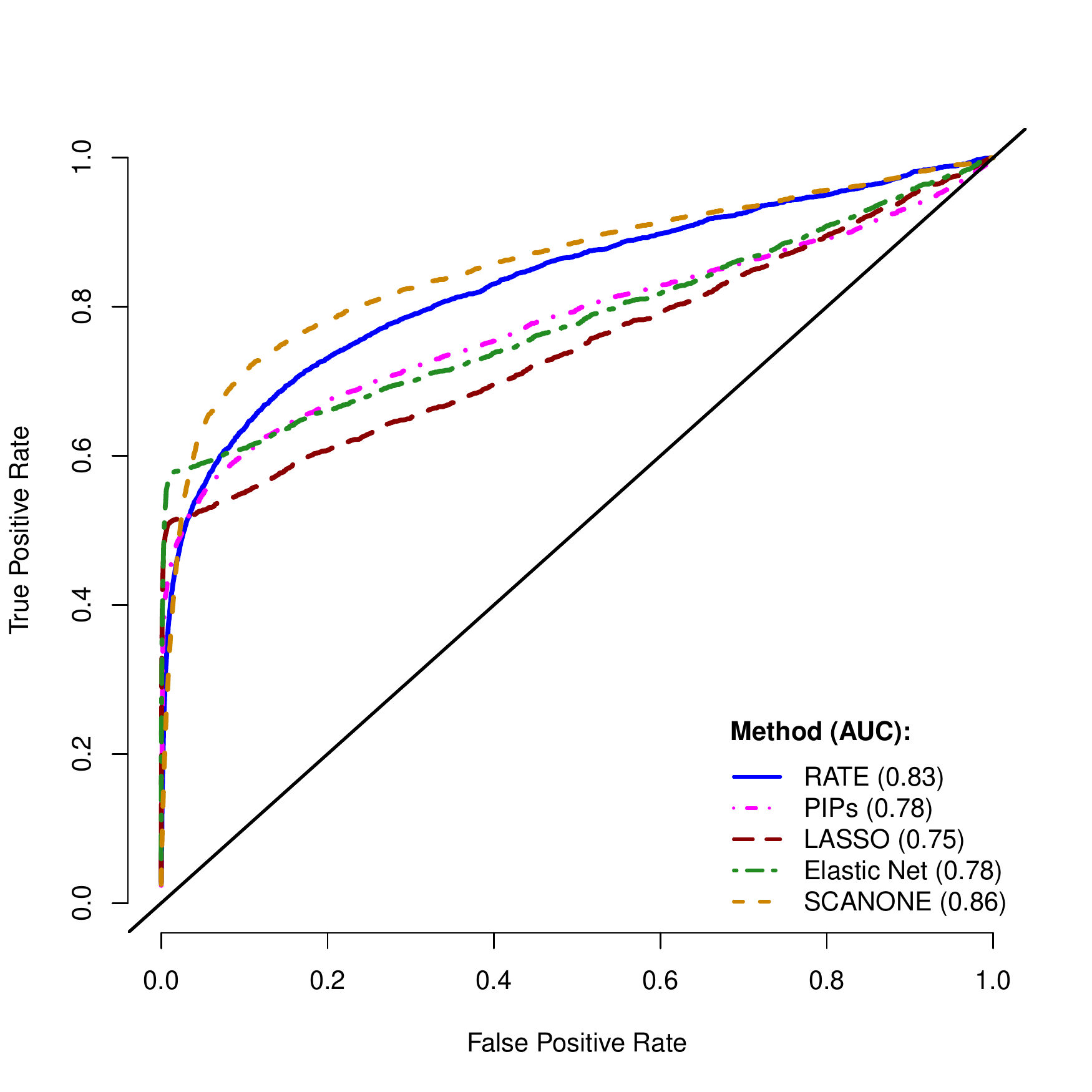}
   \label{Fig3B}
 }
 \subfigure[Scenario II ($\rho = 0.5$)]{
   \includegraphics[width = 0.48\textwidth]{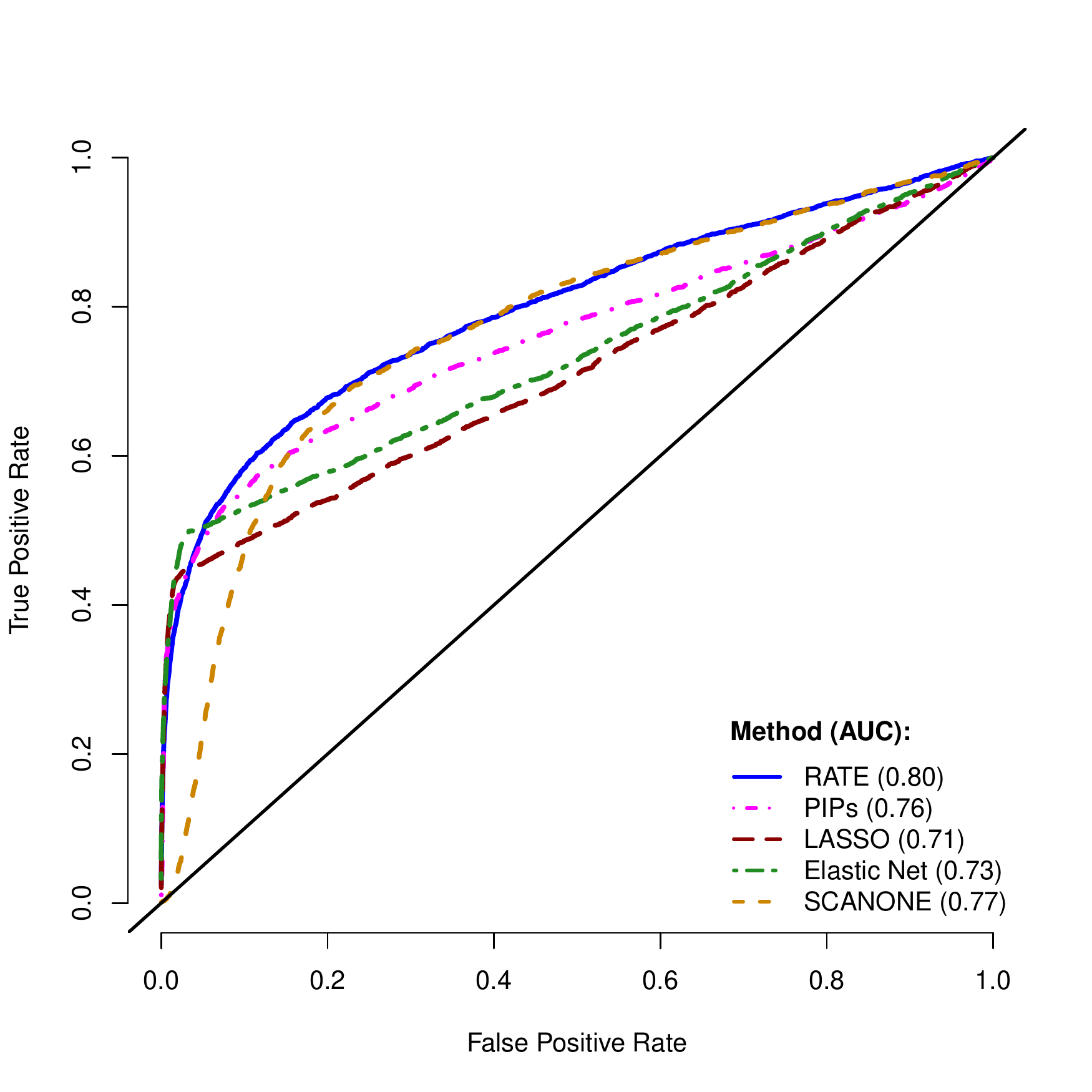}
   \label{Fig3C}
 }
 \subfigure[Scenario II ($\rho = 1$)]{
   \includegraphics[width = 0.48\textwidth]{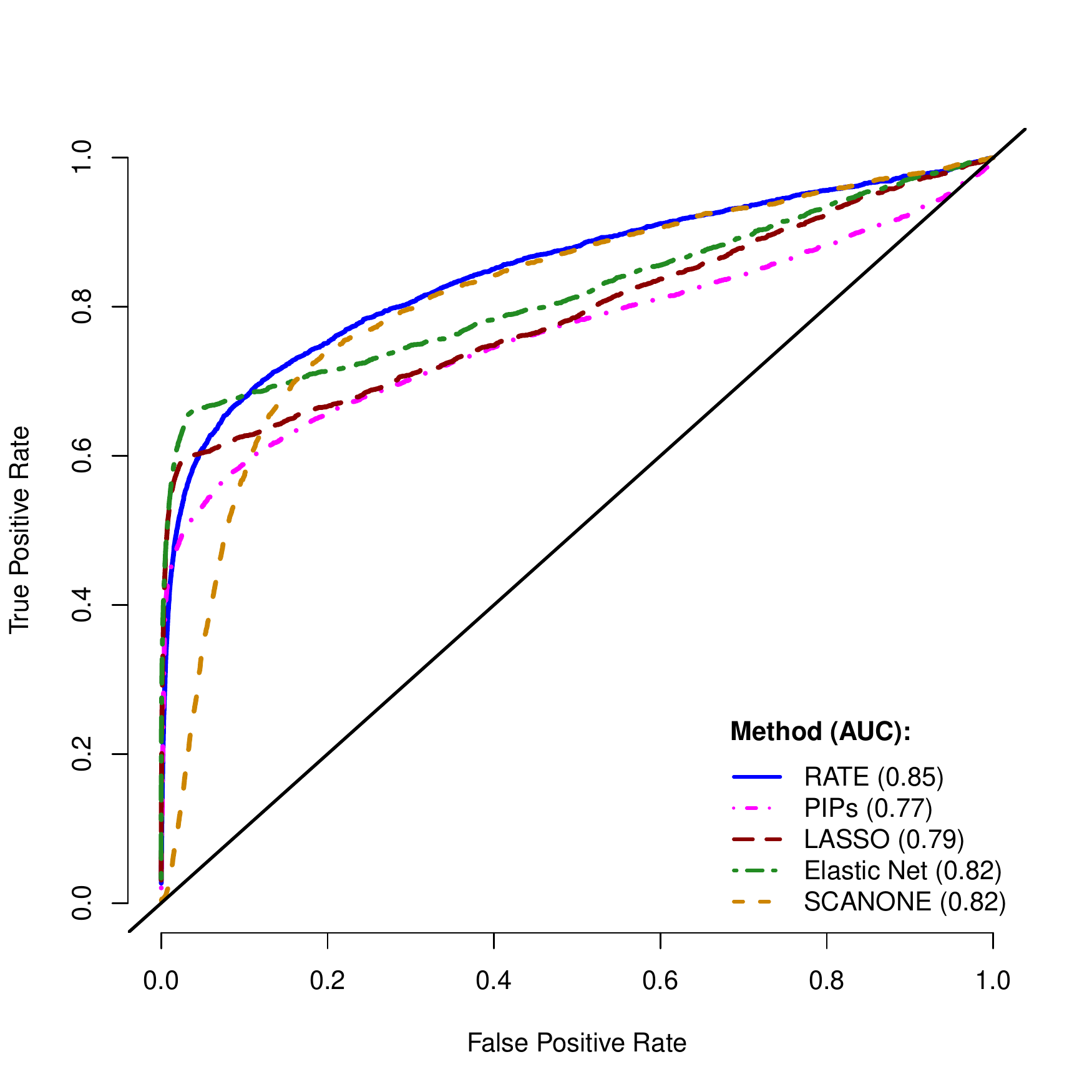}
   \label{Fig3D}
 }
\caption{\textbf{Power analysis for prioritizing genetic variants.} Phenotypes are simulated with broad-sense heritability level $\mbox{H}^2=0.3$ with control parameter $\rho = \{0.5,1\}$ in Figures (a) and (c) and Figures (b) and (d), respectively. Here, $(1-\rho)$ is used to determine the proportion of signal that is contributed by interaction effects. Compared approaches include Gaussian process regression with RATE (blue), Bayesian variable selection with a spike and slab prior (PIPs) (pink), lasso regression (red), the elastic net (green), and the SCANONE method (orange). Area under the curve (AUC) is reported to facilitate comparisons. Scenario I corresponds to phenotypic outcomes being generated via simulation model (i). Scenarios II introduces population stratification effects with simulation model (ii) by allowing the top 5 genotype PCs to make up 30\% of the phenotypic variance. Results are based on 100 replicates in each case.}
\label{Fig3}
\end{figure}

All results described in the main text are based on scenarios I and II, while results for scenario III be found in Supplementary Material. We evaluate each method's ability to effectively prioritize the causal SNPs in 100 different simulated datasets. The criteria we use compares the false positive rate (FPR) with the rate at which true variants are identified by each model (TPR). This is further quantified by assessing the area under the curve (AUC). Note that SCANONE produces p-values, lasso and the elastic net give magnitude of regression coefficients, and the Bayesian variable selection model computes posterior inclusion probabilities (PIPs). Method performance varies depending on the two factors: (a) the presence of interaction effects, and (b) additional structure due to population stratification. For example, in the first simulation scenario, all methods exhibit lower power when a proportion of the broad-sense heritability is made up of interaction effects (e.g.~Figure \ref{Fig3A}). This power increases when additive effects dominate the heritability (e.g.~Figure \ref{Fig3B}). Overall, the lasso is the worst performing method. In the cases where there are no additional population stratification effects, the SCANONE approach proved to be better method. These results are unsurprising since this scenario best suites the assumptions of this approach.

While the performance of our distributional centrality measures are comparable in the first setting, its true advantage becomes apparent when there is some underlying population structure between genotypes (i.e.~scenarios II and III). Importantly, under this type of data, the power of RATE is consistently better than its counterparts (e.g.~Figures \ref{Fig3C}, \ref{Fig3D}, and S8). From a significance threshold perspective, RATE also proves to have the best ``optimal'' selection metric. Solely considering SNPs with $\text{RATEs}>1/p$ consistently yielded more associative mapping power than observing both (a) the equivalence of the Bayesian ``median probability model'' (i.e.~$\text{PIPs}>0.5$) \cite{Barbieri:2004aa}, and (b) SCANONE p-values below the Bonferroni-corrected significance threshold (i.e.~$P<8.7\times10^{-6}$) (see Figure S9). For example, in simulation scenario II, the ``optimal'' RATE model identified 72\% and 78\% of the casual variables for $\rho = 0.5$ and $1$, respectively. This compared to 24\% and 37\% for the median probability model, and 32\% and 46\% for the multiple testing corrected SCANONE model (see Figure S9). This trend is consistent across all of the simulation settings that we consider.

Altogether, we want to stress that these simulation results are important from a model interpretation perspective. Even though methods like SCANONE effectively prioritize SNPs in certain scenarios, their significance metrics struggle to create separation between selected and non-selected markers. Therefore, if a practitioner were to choose variants satisfying some ``optimal'' genome-wide threshold, the more conservative methods will simply miss the majority of the true causal variables (i.e.~a higher count of false negatives). RATE, on the other hand, is consistently able to distinguish among the SNPs in a given set. Even in the scenarios where phenotypes are simulated without population stratification effects, RATE is more likely to deem associated variants as significant genome-wide --- just at the possible cost of slightly more false positives.

\subsection{Real Data Analysis: \textit{Arabidopsis} QTL Study}

We now apply our approach to a quantitative trait loci (QTL) association mapping study focused on the characterization of complex phenotypes in \textit{Arabidopsis thaliana}, a small flowering plant native to Eurasia. The specific dataset that we consider comes from the Versailles Arabidopsis Stock Center \cite{Loudet:2002aa} (\url{http://publiclines.versailles.inra.fr/page/33}) and has been previously used for evaluating the mapping power of other statistical methods \cite{Demetrashvili:2013aa}. More descriptively, it consists of $n=403$ F6 plants from a Bay-0 $\times$ Shahdara recombinant inbred lines (RILs) population that were genotyped for $p = 1028$ genetic markers and phenotyped for sixty-three different metabolic traits \cite{Wentzell:2007aa}. After pruning the genotypes of variants with near perfect correlation ($R \ge 0.99$), we obtained a final set of 524 markers (see Supplementary Material for details). We limit the scope of our analysis to six biochemical content measurements including: allyl, Indol-3-ylmethyl (I3M), 4-methoxy-indol-3-ylmethyl (MO4I3M), 4-methylsulfinylbutyl (MSO4), 8-methylthiooctyl (MT8), and 3-hydroxypropyl (OHP3) (see Table S2). Importantly, the goal of the original study was to highlight complex connections between gene expression and metabolite (glucosinolate) variation \cite{Wentzell:2007aa}. Here, we consider this particular case study not only because it presents a variety of quantitative traits, but also because the data contains a mixture of additive and some epistatic effects. Indeed, this dataset presents a realistic mix between the cases we previously examined for simulation scenario I.

\begin{figure}[H]
\centering
\subfigure[Gaussian Process Regression]{
   \includegraphics[width = \textwidth]{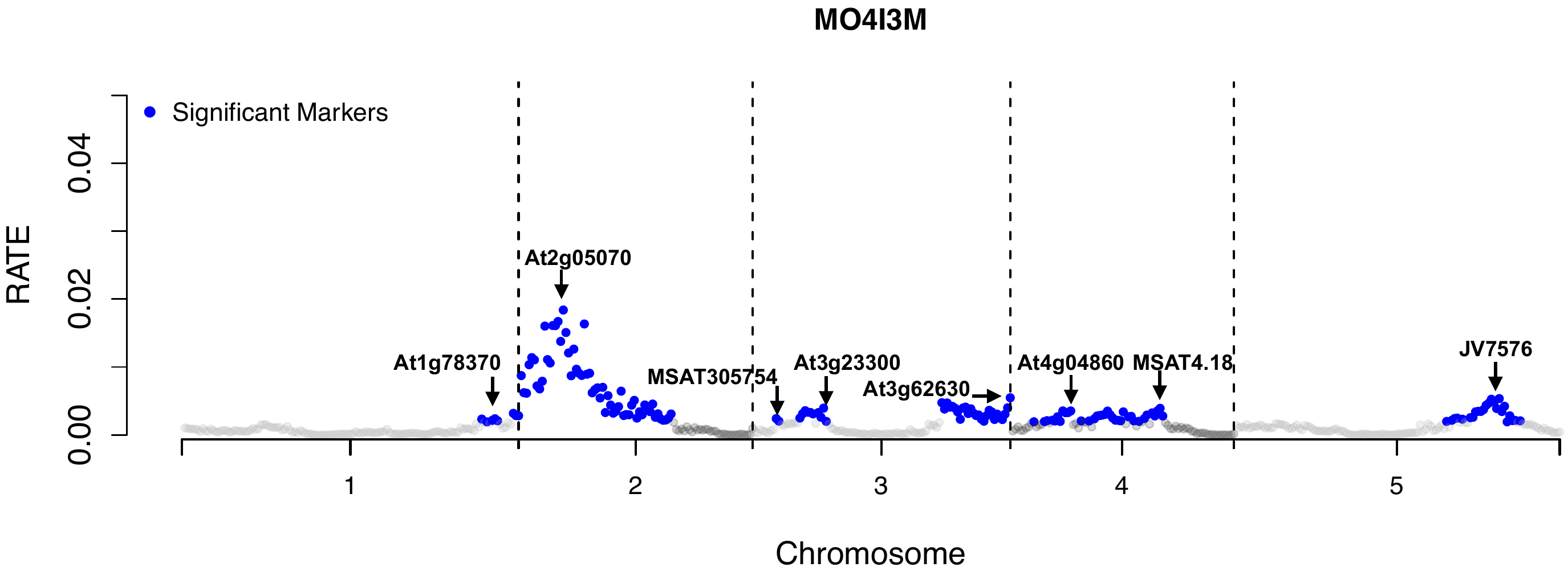}
   \label{Fig4A}
 }
  \subfigure[SCANONE]{
   \includegraphics[width = \textwidth]{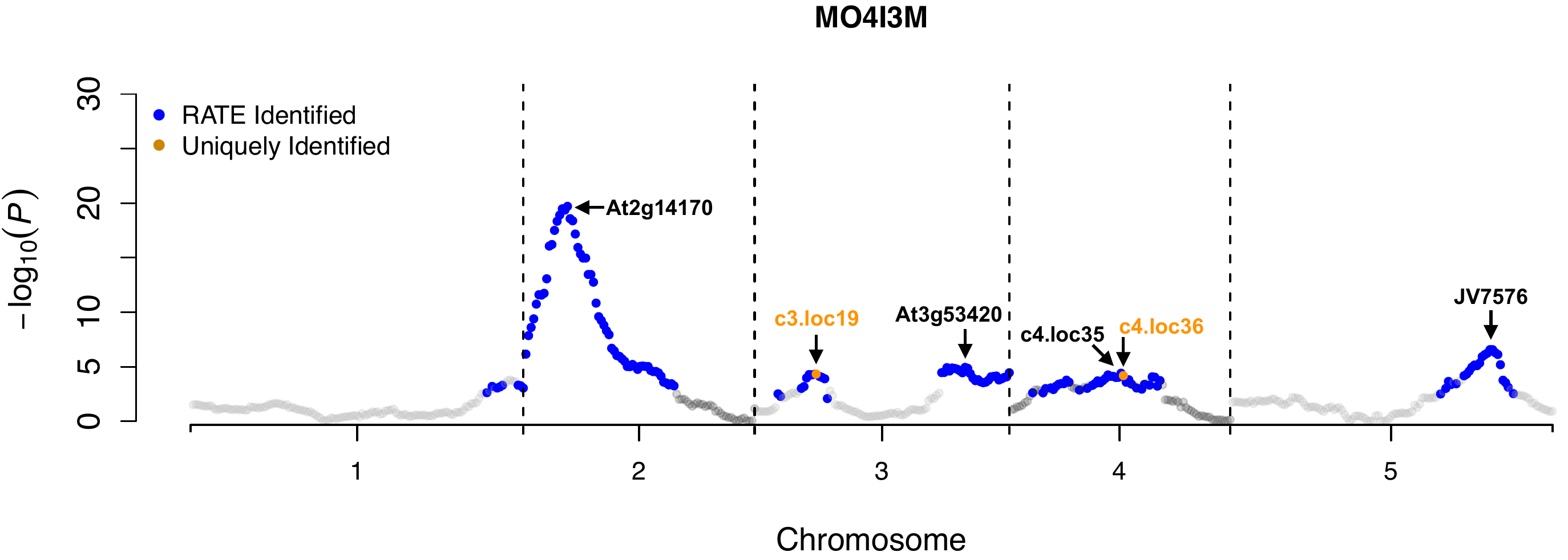}
   \label{Fig4B}
 }
\caption{\textbf{Genetic map wide scan for the 4-methoxy-indol-3-ylmethyl (MO4I3M) glucosinolate trait analyzed in \textit{Arabidopsis thaliana} QTL mapping study.} Compared methods are (a) Gaussian process regression with RATE and (b) SCANONE (orange). Significant markers are determined by $\text{RATE}(\widetilde{\beta})>1/p$ and $P<9\times10^{-5}$, respectively. The latter represents the genome-wide Bonferroni-corrected significance threshold. To ease the comparisons, points in blue represent genetic markers with significant distributional centrality measures. Markers labeled in color were not found by RATE.
}
\label{Fig4}
\end{figure}

For each metabolic trait, we provide a summary table which lists centrality measures for all gene expression polymorphisms as detected via GP regression and RATE (see Table S3). To contrast the associations identified by our nonparametric method, we also directly compared results from implementing the SCANONE approach since it proved to be the most powered of the competing methods in simulations (again see Table S3). Figures \ref{Fig4} and S10-S14 display plots of enrichment for a genome-wide scan on the six traits according to the RATE enrichment metric. These figures also show the comparative results for the standard single-variant testing approach. Referenced in all images are blue points which represent genetic markers with significant distributional centrality measures above the line of relative equivalence (i.e.~$\text{RATEs}>1/p$). In Table \ref{Tab1}, we report the number of significant markers that are identified by both methods. Once again, these are determined by markers with $\text{RATE}(\widetilde{\beta})>1/p$ and $P<9\times10^{-5}$, respectively. Again, the latter represents the genome-wide Bonferroni-corrected significance threshold. In the second part of Table \ref{Tab1}, we take the significant markers identified by each model and refit simple linear regressions with them. Here, we report $R^2$ as a way to assess which method was able to select markers that explain the greatest proportion of variance in all six traits. 

Overall, RATE consistently identified genomic locations that correspond to known members of biosynthetic pathways in \textit{Arabidopsis thaliana}. Most of these, as in the original study, were small networks of QTLs known to control biosynthetic pathways. For example, in MO4I3M, the most central loci appeared on the second chromosome and were headlined by the marker tagged At2g14170 (see Figure \ref{Fig4A}). This variant is associated with \textit{ALDH6B2} --- a gene within the \textit{Arabidopsis} genome known to catalyze enzymatic reactions in valine and pyrimidine catabolism (i.e.~destructive metabolism) \cite{Kirch:2004aa,Hou:2015aa}. Similarly, on the first chromosome, RATE featured a small group of central loci lead by At1g78370 --- which encodes a core glucosinolate biosynthesis gene \textit{GSTU20} and plays a key role in glutathione transferase activity and metabolism \cite{Wu:2016aa}. For the trait MT8 content, RATE deemed the most important region of the genome to be on the fifth chromosome (see Figure S13). Here, the marker At5g22630 had the greatest relative centrality measure. This polymorphism represents \textit{ADT5} which has recently been suggested to moonlight proteins that play an enzymatic role in biosynthesis \cite{Bross:2017aa}. This same marker is also highlighted as being moderately influential in explaining the variability in allyl content across the plants (see Figure S10). This makes sense because of the strong positive correlation between the content of these two traits.  

These validated findings from previous experimentally based studies lead us to believe that our results contain true positives. Lastly, in order to bolster confidence in the relative centrality measures identified by our nonparametric approach, we also display the correlation structure across the genotypes and phenotypes for the 403 Bay-0 $\times$ Shahdara RILs (see Figures S15 and S16). Consistent with our results, there appeared to be strong \textit{cis}-type covariances between groups of genetic markers located on the same chromosome. This underlying genetic architecture resembles data analytic situations where our approach is most powered.

In order to better explain why our nonparametric approach and the SCANONE method performed similarly in each of the six phenotypes, we use a variance component analysis to evaluate how different types of genetic effects (i.e.~linear vs.~nonlinear) contribute to the overall broad-sense heritability \cite{Zhou:2017aa} (see text in Supplementary Material for details). Briefly, we use a linear mixed model with multiple random effects to partition the phenotypic variance into three different categories: (a) an additive component, (b) a pairwise interaction component, and (c) a third order interaction component. Disregarding the contribution of random noise, we quantify the contribution of these genetic effects by estimating the proportion of heritability that is explained via their corresponding variance components. Table S4 displays these results which effectively highlights that each of the six traits are primarily dominated by additivity.

\subsection{Real Data Analysis: Heterogenous Stock of Mice GWAS}

We lastly assess RATE's association mapping ability in a more traditional GWAS setting by analyzing three quantitative traits in a heterogeneous stock of mice dataset \cite{Valdar:2006aa} from the Wellcome Trust Centre for Human Genetics (\url{http://mtweb.cs.ucl.ac.uk/mus/www/mouse/index.shtml}). This data contains $n\approx2,000$ individuals and $p\approx10,000$ SNPs with minor allele frequencies above 5\% --- with exact numbers varying slightly depending on the phenotype (see Supplementary Material for details). The three quantitative traits we consider include: body weight, percentage of CD8+ cells, and high-density lipoprotein (HDL) content. We consider this particular dataset not only because it contains a wide variety of quantitative traits, but also because the data contains related samples. Relatedness has been shown to manifest different orders of interaction effects \cite{Hemani:2013aa,Crawford:2017aa,Crawford:2017ab}, and thus this dataset also presents a realistic mix between the cases we examined in simulation scenarios II and III.

\begin{figure}[H]
\centering
\subfigure[Gaussian Process Regression]{
   \includegraphics[width = \textwidth]{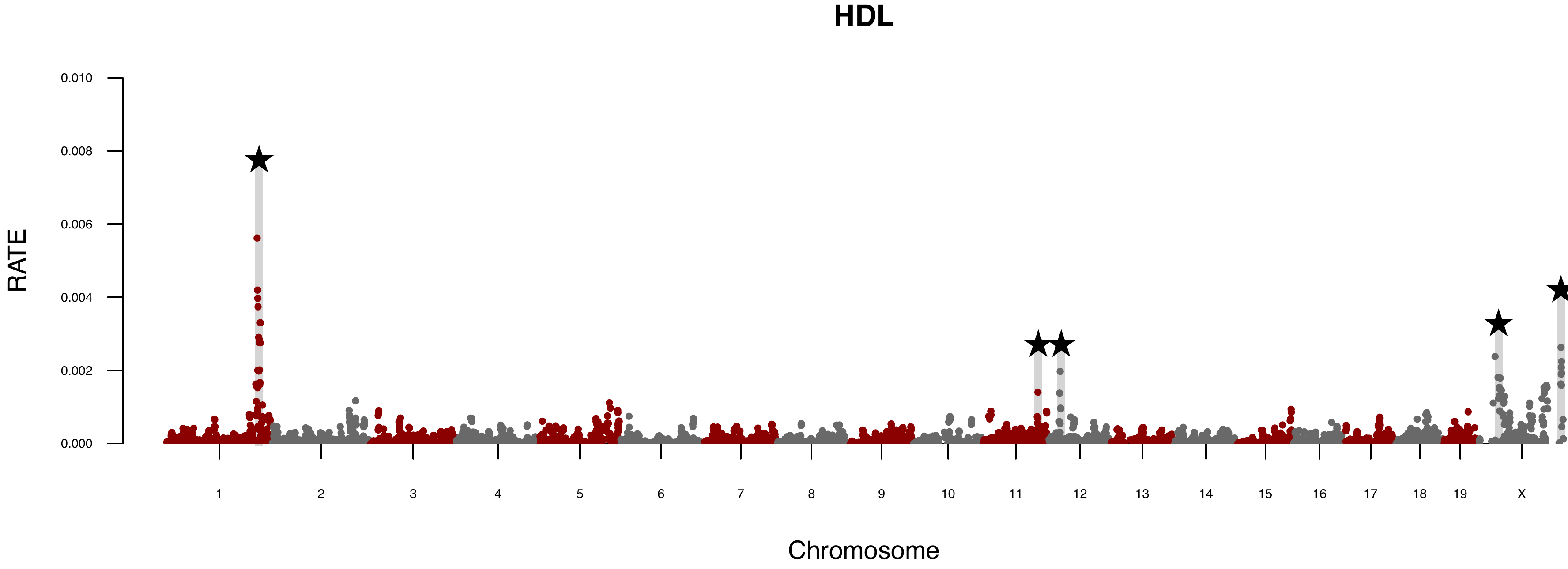}
   \label{Fig5A}
 }
 \subfigure[SCANONE]{
   \includegraphics[width = \textwidth]{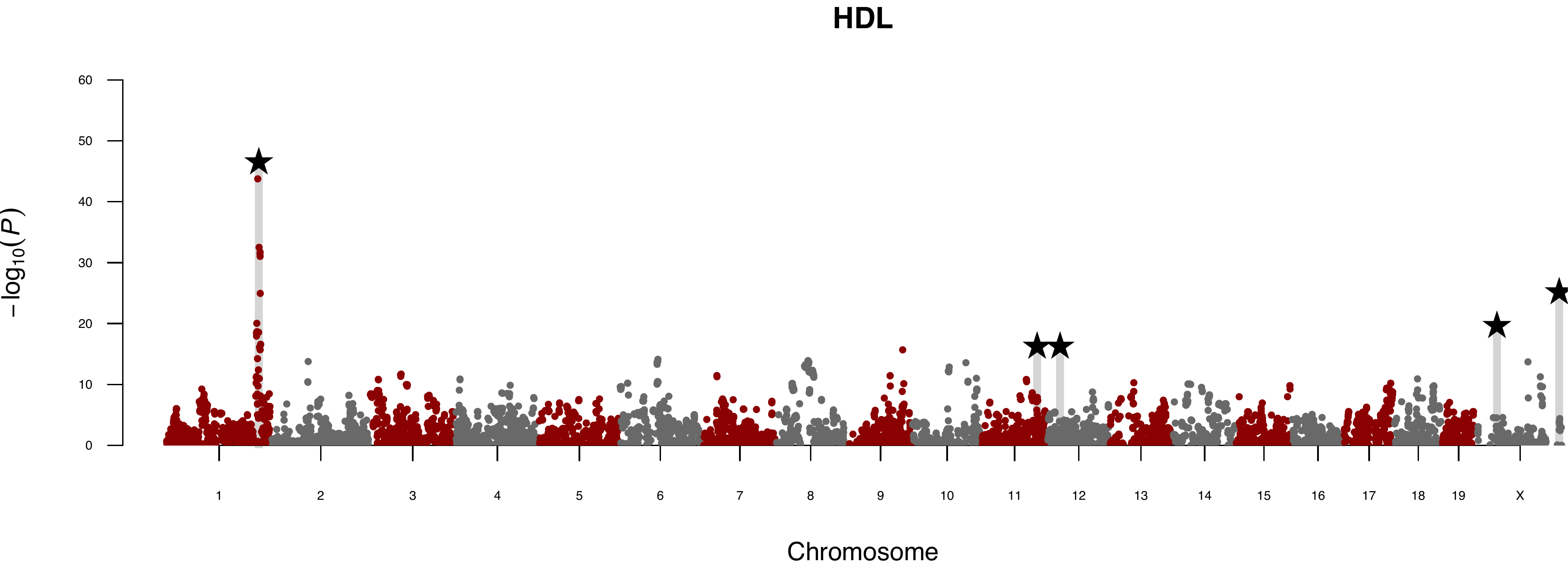}
   \label{Fig5B}
 }
\caption{\textbf{Genome-wide scan for high-density lipoprotein (HDL) content in the heterogeneous stock of mice dataset.} Figure (a) depicts the relative distributional centrality measures (RATE) of quality-control-positive SNPs plotted against their genomic positions. Gaussian process regression was used to derive these measures. Chromosomes are shown in alternating colors for clarity, with the top five most enriched regions (according to RATE) being highlighted by the star symbol. Figure (b) serves as a direct comparison and depicts results from a typical GWAS analysis using SCANONE. Here, we overlay the enriched regions detected by RATE to simplify the comparison.}
\label{Fig5}
\end{figure}

Once again, we compare the GP regression model to the single-SNP approach via SCANONE, which serves as a baseline. For each trait, we provide a summary table which lists the corresponding RATEs and p-values for all SNPs (see Table S5). Figures \ref{Fig5}, S17, and S18 then visually display this information via Manhattan plots. In these figures, chromosomes are shown in alternating colors for clarity, with the top five most enriched regions (according to RATE measures) being highlighted as a way to facilitate comparisons between the mapping approaches. 

As in the previous real data application, our nonparametric approach was able to detect multiple loci that have been previously validated as having functional associations with the traits of interests. Many of these findings were also indicated in the original study that produced this dataset \cite{Valdar:2006aa}. For example, the X chromosome is well known to majorly influence adiposity and metabolism in mice \cite{Rance:1997aa,Chen:2012aa,Chen:2013aa,Cox:2014aa}. As expected, in the body weight and HDL content traits, our approach identified significant enrichment in this genomic region --- headlined by the chromatin remodeling complex gene \textit{Smarca1} in both cases. Additionally, for the body weight phenotype, RATE also prioritized markers on chromosomes 7 and 10 as having notable associations. Previous computational studies have shown variants on both of these chromosomes to have additive effects and statistical epistatic interactions that influence mice body composition \cite{Kleyn:1996aa,Brockmann:1998aa,Diament:2003aa,Ankra-Badu:2009aa}. In this particular analysis, we attribute the selection of these loci to the nonlinear properties of the Gaussian covariance function and the nonparametric nature of the GP regression model. Similarly, for HDL content, RATE found many significant SNPs on the first, eleventh and twelfth chromosomes. The corresponding spike on chromosome 1 is a genomic location that most notably harbors the HDL driver gene \textit{Ath-1} \cite{Paigen:1987aa} (see Figure \ref{Fig5A}). Finally, for the phenotype detailing the percentage of CD8+ cells, our method identified the majority of significant SNPs to be on the seventeenth chromosome --- including those within boundary of \textit{Myof1}, a gene that has been suggested to modulate cell adhesion and motility in the immune system \cite{Kim:2006aa}. Overall, this general genomic location that has been validated to greatly determine the ratio of T-cells \cite{Yalcin:2010aa}. 

Once again, we use variance component analysis to now dissect the broad-sense heritability of these three mice traits and help better explain why there could be differences in the loci discovered by RATE and SCANONE (see Table S6). As in the previous subsection, we implement a linear mixed model to partition the overall broad-sense heritability into the same additive, second order (pairwise) interaction, and third order interaction genetic effect types. Note that, unlike in the \textit{Arabidopsis} QTL study, additive effects do not dominate the genotypic contribution in any of the three mice phenotypes that we consider --- this is particularly obvious for the trait detailing the HDL content (Figure \ref{Fig5} and Table S6). Instead, the variance components corresponding to the second and third order interactions make up the majority of the broad-sense heritability. We believe that accounting for these nonlinear relationships, as well as controlling for the relatedness between samples, allows RATE to identify loci that SCANONE misses.
 

\section{Discussion}\label{sec5}

In this paper, we proposed a new general measure for conducting variable selection in ``black box'' Bayesian methodologies. While many of these black box approaches often give notable predictive performance gains, the reasoning behind these results can be difficult to explain and interpret. Within a statistical genetics context, we discussed how the previously proposed effect size analog for nonparametric regression enables the prioritization of variants based on their marginal associations. Recognizing that one of the main sources of performance gains in black box modeling is through underlying interactions and nonlinear effects between predictor variables, we introduced our new distributional centrality measure RATE --- meant to rank genetic markers based on their influence on the joint distribution with other markers. As we demonstrated with simulation studies, our new measure can be used for feature selection, giving state-of-the-art performance even in the presence of population structure. In real QTL and GWAS data applications, RATE allowed us to uncover biologically relevant markers by simultaneously taking into account significant interactions when ranking variants based on their relative importance.

In its current form, we have focused on demonstrating RATE with a Gaussian process regression model. Although our entire illustration of the method is based on the manipulation of approximate posterior distributions in Bayesian applications, each of the innovations that we present can be applied in a frequentist setting. The effect size analog is merely a summary statistic which can be derived after fitting any model. Therefore, one could envision a frequentist setting in which parameter estimation and uncertainty is done using bootstrap, for example. In particular, this would lead to a multivariate normal-like estimator for the mean and covariance of the effect size analog. One could then proceed to compute the relative centrality measures with this distribution. The utility of our approach, from this alternative point of view, remains an open question.

RATE is not without its limitations. One particular limitation of RATE is that while it provides a measure of general association for nonparametric methods, it cannot be used to directly identify the component (i.e.~linear vs. nonlinear) that drives individual variable associations. Thus, despite being able to detect significant variants that are associated to a response in a nonlinear fashion, the RATE measure is unable to directly identify the detailed orders of interaction effects. A key part of our future work is learning how to disentangle this information. A second, and perhaps the most noticeable, limitation of RATE is that the computation of the centrality measures scales at least cubically with the number of features in the input data (see Table S7 in Supplementary Material). This is opposed to the other methods we compare in this study (e.g.~single-SNP tests) which take a fraction of the time to compute. In future work, we would like to consider the challenges of analyzing large scale studies. An example of this would be consortium-sized efforts in human-based genome-wide association studies with millions of markers and thousands of genotyped individuals \cite{Consortium:2010aa,WTCCC,Sudlow:2015aa}. In these settings, one possible immediate fix would be to use a two step procedure. In the first step, we implement a more scalable mapping method \cite{Purcell:2007aa,Lippert:2011aa,Zhou:2012aa} as a screen to select the top marginally associated markers. Then in the second step, we test for more detailed nonlinear prioritization using centrality measures. Nonetheless, new algorithms and alternative code implementations are likely needed to scale RATE up to datasets that are orders of magnitude larger in size.


\section*{Supplementary Material}

Supplementary Material is available for download at \url{http://www.lcrawlab.com/Papers/RATE_SI.pdf}.


\section*{Acknowledgements}

LC, SRF, DER, and MW would like to the Editor, Associate Editor, and two anonymous referees for their constructive comments. We would also like to thank Andrew Gelman, Elizabeth R. Hauser, Steve Oudot, Sohini Ramachandran, and Xiang Zhou for helpful conversations and suggestions. LC would also like to acknowledge the support of start up funds from Brown University. Any opinions, findings, and conclusions or recommendations expressed in this material are those of the author(s) and do not necessarily reflect the views of any of the funders. This study makes use of data generated by the Wellcome Trust Case Control Consortium (WTCCC). A full list of the investigators who contributed to the generation of the data is available from \url{www.wtccc.org.uk}. Funding for the WTCCC project was provided by the Wellcome Trust under award 076113, 085475, and 090355.


\section*{Author Contributions Statement}

LC conceived the study. LC and MW developed the methods. LC, SRF, and DER developed the algorithms. LC implemented the software and performed the analyses. All authors wrote and revised the manuscript.


\section*{Competing Financial Interests}

The authors have declared that no competing interests exist.


\newpage
\section*{Figures and Tables}

\setlength{\extrarowheight}{3pt}
\begin{table}[H]
\centering
\caption{\textbf{Comparing RATE and the SCANONE mapping approach in the \textit{Arabidopsis} QTL study.} Glucosinolate content traits include: allyl content, indol-3-ylmethyl (I3M), 4-methoxy-indol-3-ylmethyl (MO4I3M), 4-methylsulfinylbutyl (MSO4), 8-methylthiooctyl (MT8), and 3-hydroxypropyl (OHP3). Significant markers are determined by $\text{RATE}(\widetilde{\beta})>1/p$ and $P<9\times10^{-5}$, respectively. The latter represents the genome-wide Bonferroni-corrected significance threshold. Values in bold denote the best according to $R$\textsuperscript{2} when considering ``optimal'' model fit with the significant markers. The last section describes the percent overlap between the significant markers found using the two methods.}
\vspace{1em}
\begin{tabular}{c|c|cccccc}
  \hline
& & \multicolumn{6}{c}{Phenotypic Traits} \\[2pt]\hline
Category & Method & Allyl & I3M & MO4I3M & MSO4 & MT8 & OHP3\\[2pt]
\hline
\multirow{2.5}{*}{\# Sig.~Markers} & RATE & 64 & 130 & 165 & 117 & 85 & 96 \\[2pt]
& SCANONE & 61 & 75 & 99 & 100 & 71 & 98 \\[2pt]
\hline
\multirow{2.5}{*}{$R$\textsuperscript{2} of Sig.~Model} & RATE & \textbf{0.686} & \textbf{0.472} & \textbf{0.570} & \textbf{0.544} & \textbf{0.610} & \textbf{0.569} \\[2pt]
& SCANONE & 0.675 & 0.353 & 0.452 & 0.494 & 0.527 & 0.566 \\[2pt]
\hline
\% Overlap & SCANONE $\subseteq$ RATE & 97\% & 100\% & 98\% & 100\% & 100\% & 97\% \\[2pt]
\hline
\end{tabular}
\label{Tab1}
\end{table}

\clearpage
\newpage
\bibliography{RATE_Ref} 

\begin{thebibliography}{10}

\bibitem{Cotter:2011aa}
Cotter A, Keshet J, Srebro N.
\newblock Explicit approximations of the {G}aussian kernel.
\newblock arXiv. 2011;p. 1109.4603.

\bibitem{Wahba:1990aa}
Wahba G.
\newblock Splines models for observational data. vol.~59 of Series in Applied
  Mathematics.
\newblock Philadelphia, PA: SIAM; 1990.

\bibitem{Barbieri:2004aa}
Barbieri MM, Berger JO.
\newblock Optimal predictive model selection.
\newblock Ann Statist. 2004;32(3):870--897.
\newblock Available from: \url{http://projecteuclid.org/euclid.aos/1085408489}.

\bibitem{Richard:1991aa}
Richard MD, Lippmann RP.
\newblock Neural network classifiers estimate {Bayesian} \textit{a posteriori}
  probabilities.
\newblock Neural Comput. 1991;3(4):461--483.

\bibitem{Chipman:2010aa}
Chipman HA, George EI, McCulloch RE.
\newblock {BART: Bayesian additive regression trees}.
\newblock Ann Appl Stat. 2010;4(1):266--298.

\bibitem{Rasmussen}
Rasmussen CE, Williams CKI.
\newblock {G}aussian processes for machine learning.
\newblock Cambridge, MA: MIT Press; 2006.

\bibitem{Gelman:2014aa}
Gelman A, Hwang J, Vehtari A.
\newblock Understanding predictive information criteria for Bayesian models.
\newblock Stat Comput. 2014;24(6):997--1016.

\bibitem{Lin:2016aa}
Lin L, Chan C, West M.
\newblock Discriminative variable subsets in {Bayesian} classification with
  mixture models, with application in flow cytometry studies.
\newblock Biostatistics. 2016;17(1):40--53.
\newblock Available from: \url{http://dx.doi.org/10.1093/biostatistics/kxv021}.

\bibitem{lim2016estimation}
Lim C, Yu B.
\newblock Estimation stability with cross-validation (ESCV).
\newblock J Comput Graph Stat. 2016;25(2):464--492.

\bibitem{piironen2017comparison}
Piironen J, Vehtari A.
\newblock Comparison of {Bayesian} predictive methods for model selection.
\newblock Stat Comput. 2017;27(3):711--735.

\bibitem{Carvalho:2010aa}
Carvalho CM, Polson NG, Scott JG.
\newblock The horseshoe estimator for sparse signals.
\newblock Biometrika. 2010;97(2):465--480.

\bibitem{Zhang:2007aa}
Zhang Y, Liu JS.
\newblock {B}ayesian inference of epistatic interactions in case-control
  studies.
\newblock Nat Genet. 2007;39(9):1167--1173.
\newblock Available from: \url{http://dx.doi.org/10.1038/ng2110}.

\bibitem{Phillips:2008aa}
Phillips PC.
\newblock Epistasis---the essential role of gene interactions in the structure
  and evolution of genetic systems.
\newblock Nat Rev Genet. 2008;9(11):855--867.
\newblock Available from:
  \url{http://www.ncbi.nlm.nih.gov/pmc/articles/PMC2689140/}.

\bibitem{Wan:2010aa}
Wan X, Yang C, Yang Q, Xue H, Fan X, Tang NL, et~al.
\newblock BOOST: a fast approach to detecting gene-gene interactions in
  genome-wide case-control studies.
\newblock Am J Hum Genet. 2010;87(3):325--340.

\bibitem{Zhang:2010aa}
Zhang X, Huang S, Zou F, Wang W.
\newblock TEAM: efficient two-locus epistasis tests in human genome-wide
  association study.
\newblock Bioinformatics. 2010;26(12):i217--i227.
\newblock Available from:
  \url{http://www.ncbi.nlm.nih.gov/pmc/articles/PMC2881371/}.

\bibitem{Prabhu:2012aa}
Prabhu S, Pe'er I.
\newblock Ultrafast genome-wide scan for {SNP}-{SNP} interactions in common
  complex disease.
\newblock Genome Res. 2012;22(11):2230--2240.
\newblock Available from:
  \url{http://www.ncbi.nlm.nih.gov/pmc/articles/PMC3483552/}.

\bibitem{Mackay:2014aa}
Mackay TFC.
\newblock Epistasis and quantitative traits: using model organisms to study
  gene-gene interactions.
\newblock Nat Rev Genet. 2014;15(1):22--33.
\newblock Available from: \url{http://dx.doi.org/10.1038/nrg3627}.

\bibitem{Crawford:2017ab}
Crawford L, Zeng P, Mukherjee S, Zhou X.
\newblock Detecting epistasis with the marginal epistasis test in genetic
  mapping studies of quantitative traits.
\newblock PLoS Genet. 2017;13(7):e1006869--.
\newblock Available from: \url{https://doi.org/10.1371/journal.pgen.1006869}.

\bibitem{Crawford:2018aa}
Crawford L, Zhou X.
\newblock Genome-wide Marginal Epistatic Association Mapping in Case-Control
  Studies.
\newblock bioRxiv. 2018;p. 374983.
\newblock Available from:
  \url{http://biorxiv.org/content/early/2018/07/23/374983.abstract}.

\bibitem{Horn:2011aa}
Horn T, Sandmann T, Fischer B, Axelsson E, Huber W, Boutros M.
\newblock Mapping of signaling networks through synthetic genetic interaction
  analysis by {RNAi}.
\newblock Nat Meth. 2011;8(4):341--346.
\newblock Available from: \url{http://dx.doi.org/10.1038/nmeth.1581}.

\bibitem{Howard:2014aa}
Howard R, Carriquiry AL, Beavis WD.
\newblock Parametric and nonparametric statistical methods for genomic
  selection of traits with additive and epistatic genetic architectures.
\newblock G3 (Bethesda). 2014;4(6):1027--1046.

\bibitem{Hill:2008aa}
Hill WG, Goddard ME, Visscher PM.
\newblock Data and theory point to mainly additive genetic variance for complex
  traits.
\newblock PLoS Genet. 2008;4(2):e1000008.
\newblock Available from:
  \url{http://dx.doi.org/10.1371%2Fjournal.pgen.1000008}.

\bibitem{Hemani:2014aa}
Hemani G, Shakhbazov K, Westra HJ, Esko T, Henders AK, McRae AF, et~al.
\newblock Detection and replication of epistasis influencing transcription in
  humans.
\newblock Nature. 2014;508(7495):249--253.
\newblock Available from: \url{http://dx.doi.org/10.1038/nature13005}.

\bibitem{Wood:2014aa}
Wood AR, Tuke MA, Nalls MA, Hernandez DG, Bandinelli S, Singleton AB, et~al.
\newblock Another explanation for apparent epistasis.
\newblock Nature. 2014;514(7520):E3--E5.

\bibitem{Wang:2010aa}
Wang X, Elston RC, Zhu X.
\newblock Statistical interaction in human genetics: how should we model it if
  we are looking for biological interaction?
\newblock Nat Rev Genet. 2010;12:74.
\newblock Available from: \url{http://dx.doi.org/10.1038/nrg2579-c2}.

\bibitem{Wang:2011aa}
Wang X, Elston RC, Zhu X.
\newblock The meaning of interaction.
\newblock Hum Hered. 2011;70(4):269--277.
\newblock Available from:
  \url{http://www.ncbi.nlm.nih.gov/pmc/articles/PMC3025890/}.

\bibitem{Crawford:2017aa}
Crawford L, Wood KC, Zhou X, Mukherjee S.
\newblock Bayesian approximate kernel regression with variable selection.
\newblock J Am Stat Assoc. 2018;Available from:
  \url{https://doi.org/10.1080/01621459.2017.1361830}.

\bibitem{MJ:2011aa}
Zhang Z, Dai G, Jordan MI.
\newblock {B}ayesian generalized kernel mixed models.
\newblock J Mach Learn Res. 2011;12:111--139.

\bibitem{Merc:1909}
Mercer J.
\newblock Functions of positive and negative type and their connection with the
  theory of integral equations.
\newblock Philos Trans Royal Soc A. 1909;209:415--446.

\bibitem{Wolpert:2007aa}
Pillai NS, Wu Q, Liang F, Mukherjee S, Wolpert R.
\newblock Characterizing the function space for {B}ayesian kernel models.
\newblock J Mach Learn Res. 2007;8:1769--1797.

\bibitem{Scholkopf:2001aa}
Sch{\"o}lkopf B, Herbrich R, Smola AJ.
\newblock A generalized representer theorem.
\newblock In: Proceedings of the 14th Annual Conference on Computational
  Learning Theory and and 5th European Conference on Computational Learning
  Theory. London, UK, UK: Springer-Verlag; 2001. p. 416--426.
\newblock Available from:
  \url{http://dl.acm.org/citation.cfm?id=648300.755324}.

\bibitem{Kolmogorov:1960aa}
Kolmogorov AN, Rozanov YA.
\newblock On strong mixing conditions for stationary {Gaussian} processes.
\newblock Theory Probab Its Appl. 1960;5(2):204--208.

\bibitem{Lippert:2011aa}
Lippert C, Listgarten J, Liu Y, Kadie CM, Davidson RI, Heckerman D.
\newblock {FaST} linear mixed models for genome-wide association studies.
\newblock Nat Meth. 2011;8(10):833--835.
\newblock Available from: \url{http://dx.doi.org/10.1038/nmeth.1681}.

\bibitem{Zhou:2012aa}
Zhou X, Stephens M.
\newblock Genome-wide efficient mixed-model analysis for association studies.
\newblock Nat Genet. 2012;44(7):821--825.

\bibitem{Kang:2010aa}
Kang HM, Sul JH, Service SK, Zaitlen NA, Kong Sy, Freimer NB, et~al.
\newblock Variance component model to account for sample structure in
  genome-wide association studies.
\newblock Nat Genet. 2010;42(4):348--354.
\newblock Available from: \url{http://dx.doi.org/10.1038/ng.548}.

\bibitem{Wu:2011}
Wu MC, Lee S, Cai T, Li Y, Boehnke M, Lin X.
\newblock Rare-variant association testing for sequencing data with the
  sequence kernel association test.
\newblock Am J Hum Genet. 2011;89(1):82--93.
\newblock Available from:
  \url{http://www.ncbi.nlm.nih.gov/pmc/articles/PMC3135811/}.

\bibitem{Yang:2014aa}
Yang J, Zaitlen NA, Goddard ME, Visscher PM, Price AL.
\newblock Advantages and pitfalls in the application of mixed-model association
  methods.
\newblock Nat Genet. 2014;46(2):100--106.
\newblock Available from: \url{http://dx.doi.org/10.1038/ng.2876}.

\bibitem{Zhou:2014aa}
Zhou X, Stephens M.
\newblock Efficient multivariate linear mixed model algorithms for genomewide
  association studies.
\newblock Nat Meth. 2014;11(4):407--409.

\bibitem{Campos:2009aa}
de~los Campos G, Naya H, Gianola D, Crossa J, Legarra A, Manfredi E, et~al.
\newblock Predicting quantitative traits with regression models for dense
  molecular markers and pedigree.
\newblock Genetics. 2009;182(1):375--385.
\newblock Available from:
  \url{http://www.genetics.org/content/182/1/375.abstract}.

\bibitem{Campos:2010aa}
de~los Campos G, Gianola D, Rosa GJM, Weigel KA, Crossa J.
\newblock Semi-parametric genomic-enabled prediction of genetic values using
  reproducing kernel {H}ilbert spaces methods.
\newblock Genet Res (Camb). 2010;92(4):295--308.

\bibitem{Shi:2012aa}
Shi JQ, Wang B, Will EJ, West RM.
\newblock Mixed-effects Gaussian process functional regression models with
  application to dose--response curve prediction.
\newblock Stat Med. 2012;31(26):3165--3177.
\newblock Available from: \url{https://doi.org/10.1002/sim.4502}.

\bibitem{Weissbrod:2016aa}
Weissbrod O, Geiger D, Rosset S.
\newblock Multikernel linear mixed models for complex phenotype prediction.
\newblock Genome Res. 2016;26(7):969--979.
\newblock Available from:
  \url{http://genome.cshlp.org/content/26/7/969.abstract}.

\bibitem{Cuevas:2017aa}
Cuevas J, Crossa J, Montesinos-L{\'o}pez OA, Burgue{\~n}o J,
  P{\'e}rez-Rodr{\'\i}guez P, de~los Campos G.
\newblock Bayesian genomic prediction with genotype × environment interaction
  kernel models.
\newblock G3 (Bethesda). 2017;7(1):41--53.
\newblock Available from:
  \url{http://www.ncbi.nlm.nih.gov/pmc/articles/PMC5217122/}.

\bibitem{Chaudhuri:aa}
Chaudhuri A, Kakde D, Sadek C, Gonzalez L, Kong S.
\newblock The mean and median criterion for automatic kernel bandwidth
  selection for support vector data description.
\newblock arXiv. 2017;p. 1708.05106.

\bibitem{Liang:2008aa}
Liang F, Paulo R, Molina G, Clyde MA, Berger JO.
\newblock Mixtures of \textit{g}-priors for {B}ayesian variable selection.
\newblock J Am Stat Assoc. 2008;103(481):410--423.

\bibitem{Drineas:2005}
Drineas P, Mahoney MW.
\newblock On the {N}ystr\"{o}m method for approximating a {Gram} matrix for
  improved kernel-based learning.
\newblock J Mach Learn Res. 2005;6:2153--2175.
\newblock Available from:
  \url{http://dl.acm.org/citation.cfm?id=1046920.1194916}.

\bibitem{Recht:2007aa}
Rahimi A, Recht B.
\newblock Random features for large-scale kernel machines.
\newblock Adv Neural Inf Process Syst. 2007;3(4):5.

\bibitem{fasshauer2016kernel}
Fasshauer G, McCourt M.
\newblock Kernel-based approximation methods using matlab.
\newblock World Scientific; 2016.

\bibitem{Jiang:2015aa}
Jiang Y, Reif JC.
\newblock Modeling epistasis in genomic selection.
\newblock Genetics. 2015;201:759--768.

\bibitem{Stephens_Nature}
Stephens M, Balding DJ.
\newblock Bayesian statistical methods for genetic association studies.
\newblock Nat Rev Genet. 2009;10:681--690.

\bibitem{goutis1998model}
Goutis C, Robert CP.
\newblock {Model choice in generalised linear models: a Bayesian approach via
  Kullback-Leibler projections}.
\newblock Biometrika. 1998;85(1):29--37.

\bibitem{Smith:2006aa}
Smith A, Naik PA, Tsai CL.
\newblock Markov-switching model selection using {Kullback--Leibler}
  divergence.
\newblock J Econom. 2006;134(2):553--577.

\bibitem{Woo:aa}
Woo JH, Shimoni Y, Yang WS, Subramaniam P, Iyer A, Nicoletti P, et~al.
\newblock Elucidating compound mechanism of action by network perturbation
  analysis.
\newblock Cell. 2015;162(2):441--451.
\newblock Available from: \url{http://dx.doi.org/10.1016/j.cell.2015.05.056}.

\bibitem{Tan:2017aa}
Tan S, Caruana R, Hooker G, Lou Y.
\newblock Detecting bias in black-box models using transparent model
  distillation.
\newblock arXiv. 2017;p. 1710.06169.

\bibitem{piironen2016projection}
Piironen J, Vehtari A.
\newblock Projection predictive model selection for {Gaussian} processes.
\newblock In: IEEE International Workshop on Machine Learning for Signal
  Processing. IEEE; 2016. p. 1--6.

\bibitem{Alaa:2017aa}
Alaa AM, van~der Schaar M.
\newblock Bayesian nonparametric causal inference: information rates and
  learning algorithms.
\newblock arXiv. 2017;p. 1712.08914.

\bibitem{Mathai:1992aa}
Mathai AM, Provost SB.
\newblock Quadratic forms in random variables.
\newblock Taylor \& Francis; 1992.
\newblock Available from: \url{https://books.google.com/books?id=tFOqQgAACAAJ}.

\bibitem{GruberWest2016BA}
Gruber LF, West M.
\newblock {GPU}-accelerated {B}ayesian learning in simultaneous graphical
  dynamic linear models.
\newblock Bayesian Anal. 2016;11:125--149.
\newblock Available from: \url{http://projecteuclid.org/euclid.ba/1425304898}.

\bibitem{GruberWest2017ECOSTA}
Gruber LF, West M.
\newblock Bayesian forecasting and scalable multivariate volatility analysis
  using simultaneous graphical dynamic linear models.
\newblock Econometrics and Statistics. 2017;3:3--22.

\bibitem{Carvalho:2007aa}
Carvalho CM, West M.
\newblock Dynamic matrix-variate graphical models.
\newblock Bayesian Anal. 2007;2(1):69--97.
\newblock Available from:
  \url{https://projecteuclid.org:443/euclid.ba/1340390064}.

\bibitem{Zeng:2017aa}
Zeng P, Zhou X.
\newblock {Non-parametric genetic prediction of complex traits with latent
  Dirichlet process regression models}.
\newblock Nat Comm. 2017;8(1):456.

\bibitem{WTCCC}
{The Wellcome Trust Case Control Consortium}.
\newblock Genome-wide association study of 14,000 cases of seven common
  diseases and 3,000 shared controls.
\newblock Nature. 2007;447(7145):661--678.
\newblock Available from: \url{http://dx.doi.org/10.1038/nature05911}.

\bibitem{Yandell:2007aa}
Yandell BS, Mehta T, Banerjee S, Shriner D, Venkataraman R, Moon JY, et~al.
\newblock {R/qtlbim: QTL with Bayesian interval mapping in experimental
  crosses}.
\newblock Bioinformatics. 2007;23(5):641--643.
\newblock Available from:
  \url{http://www.ncbi.nlm.nih.gov/pmc/articles/PMC4995770/}.

\bibitem{Waldmann:2013aa}
Waldmann P, M{\'e}sz{\'a}ros G, Gredler B, F{\"u}rst C, S{\"o}lkner J.
\newblock Evaluation of the lasso and the elastic net in genome-wide
  association studies.
\newblock Front Genet. 2013;4:270.
\newblock Available from:
  \url{https://www.frontiersin.org/article/10.3389/fgene.2013.00270}.

\bibitem{Guan:2011aa}
Guan Y, Stephens M.
\newblock Bayesian variable selection regression for genome-wide association
  studies and other large-scale problems.
\newblock Ann Appl Stat. 2011;5(3):1780--1815.
\newblock Available from:
  \url{https://projecteuclid.org:443/euclid.aoas/1318514285}.

\bibitem{Loudet:2002aa}
Loudet O, Chaillou S, Camilleri C, Bouchez D, Daniel-Vedele F.
\newblock Bay-0 $\times$ {Shahdara} recombinant inbred line population: a
  powerful tool for the genetic dissection of complex traits in {Arabidopsis}.
\newblock Theor Appl Genet. 2002;104(6):1173--1184.

\bibitem{Demetrashvili:2013aa}
Demetrashvili N, Van~den Heuvel E, Wit E.
\newblock {Probability genotype imputation method and integrated weighted lasso
  for QTL identification}.
\newblock BMC Genet. 2013;14:125.

\bibitem{Wentzell:2007aa}
Wentzell AM, Rowe HC, Hansen BG, Ticconi C, Halkier BA, Kliebenstein DJ.
\newblock Linking metabolic {QTLs} with network and {cis-eQTLs} controlling
  biosynthetic pathways.
\newblock PLoS Genet. 2007;3(9):e162--.
\newblock Available from: \url{https://doi.org/10.1371/journal.pgen.0030162}.

\bibitem{Kirch:2004aa}
Kirch HH, Bartels D, Wei Y, Schnable PS, Wood AJ.
\newblock {The ALDH gene superfamily of Arabidopsis}.
\newblock Trends Plant Sci. 2004;9(8):371--377.
\newblock Available from:
  \url{http://www.sciencedirect.com/science/article/pii/S1360138504001529}.

\bibitem{Hou:2015aa}
Hou Q, Bartels D.
\newblock {Comparative study of the aldehyde dehydrogenase (ALDH) gene
  superfamily in the glycophyte \textit{Arabidopsis thaliana} and
  \textit{Eutrema} halophytes}.
\newblock Ann Bot. 2015;115(3):465--479.
\newblock Available from:
  \url{http://www.ncbi.nlm.nih.gov/pmc/articles/PMC4332599/}.

\bibitem{Wu:2016aa}
Wu J, Zhao Q, Yang Q, Liu H, Li Q, Yi X, et~al.
\newblock {Comparative transcriptomic analysis uncovers the complex genetic
  network for resistance to \textit{Sclerotinia sclerotiorum} in
  \textit{Brassica napus}}.
\newblock Sci Rep. 2016;6:19007 EP --.
\newblock Available from: \url{http://dx.doi.org/10.1038/srep19007}.

\bibitem{Bross:2017aa}
Bross CD, Howes TR, Abolhassani~Rad S, Kljakic O, Kohalmi SE.
\newblock {Subcellular localization of \textit{Arabidopsis} arogenate
  dehydratases suggests novel and non-enzymatic roles}.
\newblock J Exp Bot. 2017;68(7):1425--1440.
\newblock Available from: \url{http://dx.doi.org/10.1093/jxb/erx024}.

\bibitem{Zhou:2017aa}
Zhou X.
\newblock A unified framework for variance component estimation with summary
  statistics in genome-wide association studies.
\newblock Ann Appl Stat. 2017;11(4):2027--2051.
\newblock Available from:
  \url{https://projecteuclid.org:443/euclid.aoas/1514430276}.

\bibitem{Valdar:2006aa}
Valdar W, Solberg LC, Gauguier D, Burnett S, Klenerman P, Cookson WO, et~al.
\newblock Genome-wide genetic association of complex traits in heterogeneous
  stock mice.
\newblock Nat Genet. 2006;38(8):879--887.
\newblock Available from: \url{http://dx.doi.org/10.1038/ng1840}.

\bibitem{Hemani:2013aa}
Hemani G, Knott S, Haley C.
\newblock An evolutionary perspective on epistasis and the missing
  heritability.
\newblock PLoS Genet. 2013;9(2):e1003295.
\newblock Available from: \url{http://dx.doi.org/10.1371}.

\bibitem{Rance:1997aa}
Rance KA, Hill WG, Keightley PD.
\newblock Mapping quantitative trait loci for body weight on the X chromosome
  in mice. I. Analysis of a reciprocal F2 population.
\newblock Genet Res. 1997 Oct;70(2):117--124.

\bibitem{Chen:2012aa}
Chen X, McClusky R, Chen J, Beaven SW, Tontonoz P, Arnold AP, et~al.
\newblock The number of X chromosomes causes sex differences in adiposity in
  mice.
\newblock PLoS Genet. 2012;8(5):e1002709--.
\newblock Available from: \url{https://doi.org/10.1371/journal.pgen.1002709}.

\bibitem{Chen:2013aa}
Chen X, McClusky R, Itoh Y, Reue K, Arnold AP.
\newblock X and Y chromosome complement influence adiposity and metabolism in
  mice.
\newblock Endocrinology. 2013;154(3):1092--1104.

\bibitem{Cox:2014aa}
Cox KH, Bonthuis PJ, Rissman EF.
\newblock Mouse model systems to study sex chromosome genes and behavior:
  relevance to humans.
\newblock Front Neuroendocrinol. 2014;35(4):405--419.

\bibitem{Kleyn:1996aa}
Kleyn PW, Fan W, Kovats SG, Lee JJ, Pulido JC, Wu Y, et~al.
\newblock Identification and characterization of the mouse obesity gene tubby:
  a member of a novel gene family.
\newblock Cell. 1996;85(2):281--290.
\newblock Available from:
  \url{http://www.sciencedirect.com/science/article/pii/S0092867400811046}.

\bibitem{Brockmann:1998aa}
Brockmann GA, Haley CS, Renne U, Knott SA, Schwerin M.
\newblock Quantitative trait loci affecting body weight and fatness from a
  mouse line selected for extreme high growth.
\newblock Genetics. 1998;150(1):369--381.
\newblock Available from:
  \url{http://www.genetics.org/content/150/1/369.abstract}.

\bibitem{Diament:2003aa}
Diament AL, Warden CH.
\newblock Multiple linked mouse chromosome 7 loci influence body fat mass.
\newblock Int J Obes. 2003;28:199 EP --.
\newblock Available from: \url{http://dx.doi.org/10.1038/sj.ijo.0802516}.

\bibitem{Ankra-Badu:2009aa}
Ankra-Badu GA, Pomp D, Shriner D, Allison DB, Yi N.
\newblock Genetic influences on growth and body composition in mice: multilocus
  interactions.
\newblock Int J Obes (Lond). 2009;33(1):89--95.

\bibitem{Paigen:1987aa}
Paigen B, Mitchell D, Reue K, Morrow A, Lusis AJ, LeBoeuf RC.
\newblock Ath-1, a gene determining atherosclerosis susceptibility and high
  density lipoprotein levels in mice.
\newblock Proceedings of the National Academy of Sciences of the United States
  of America. 1987;84(11):3763--3767.
\newblock Available from:
  \url{http://www.ncbi.nlm.nih.gov/pmc/articles/PMC304956/}.

\bibitem{Kim:2006aa}
Kim SV, Mehal WZ, Dong X, Heinrich V, Pypaert M, Mellman I, et~al.
\newblock Modulation of cell adhesion and motility in the immune system by
  Myo1f.
\newblock Science. 2006;314(5796):136--139.

\bibitem{Yalcin:2010aa}
Yalcin B, Nicod J, Bhomra A, Davidson S, Cleak J, Farinelli L, et~al.
\newblock Commercially available outbred mice for genome-wide association
  studies.
\newblock PLoS Genet. 2010;6(9):e1001085--.
\newblock Available from: \url{https://doi.org/10.1371/journal.pgen.1001085}.

\bibitem{Consortium:2010aa}
{The 1000 Genomes Project Consortium}.
\newblock A map of human genome variation from population-scale sequencing.
\newblock Nature. 2010;467:1061--1073.
\newblock Available from: \url{http://dx.doi.org/10.1038/nature09534}.

\bibitem{Sudlow:2015aa}
Sudlow C, Gallacher J, Allen N, Beral V, Burton P, Danesh J, et~al.
\newblock UK Biobank: an open access resource for identifying the causes of a
  wide range of complex diseases of middle and old age.
\newblock PLoS Med. 2015;12(3):e1001779--.
\newblock Available from: \url{https://doi.org/10.1371/journal.pmed.1001779}.

\bibitem{Purcell:2007aa}
Purcell S, Neale B, Todd-Brown K, Thomas L, Ferreira MAR, Bender D, et~al.
\newblock PLINK: a tool set for whole-genome association and population-based
  linkage analyses.
\newblock Am J Hum Genet. 2007;81(3):559--575.
\newblock Available from:
  \url{//www.sciencedirect.com/science/article/pii/S0002929707613524}.

\end{thebibliography}


\end{document}